  \RequirePackage{fix-cm}
  \documentclass[twocolumn]{svjour3}          
  \smartqed  
  \usepackage{graphicx}
  \usepackage[authoryear]{natbib}
  
  \usepackage{tabu,multirow}
  \usepackage[leftcaption]{sidecap}
  \usepackage{amsmath}
  \usepackage{amssymb}
  \usepackage{color}
  \usepackage{enumitem}
  \usepackage{verbatim}
  \usepackage[colorlinks,citecolor=blue]{hyperref}
  
  \usepackage{graphicx}
  \usepackage{subfigure}
  \usepackage{float}  
  \usepackage{textcomp}
  
  \usepackage{multicol}
  \usepackage{multirow}
  
  \usepackage{eso-pic}
  \usepackage{xspace}
  \makeatletter
  \DeclareRobustCommand\onedot{\futurelet\@let@token\@onedot}
  \def\@onedot{\ifx\@let@token.\else.\null\fi\xspace}
  
  \def\eg{\emph{e.g}\onedot} 
  \def\ie{\emph{i.e}\onedot}

  \makeatother
  
  \usepackage{mathtools}
  \DeclarePairedDelimiter\ceil{\lceil}{\rceil}
  \DeclarePairedDelimiter\floor{\lfloor}{\rfloor}

\begin{document}
  \sloppy
  
  \title{Learning JPEG Compression Artifacts for Image Manipulation Detection and Localization
  }
  
  
  
  \author{Myung-Joon Kwon         \and
          Seung-Hun Nam \and
          In-Jae Yu \and
          Heung-Kyu Lee \and
          Changick Kim 
  }
  
  
  \institute{Myung-Joon Kwon \at
                School of Electrical Engineering, Korea Advanced Institute of Science and Technology (KAIST), Daejeon, South Korea \\
                \email{mjkwon2021@gmail.com}          \\
                ORCID: 0000-0002-9784-8440
             \and
             Seung-Hun Nam \at
             NAVER WEBTOON AI, Seongnam, South Korea \\
             \email{shnam1520@gmail.com} \\
             ORCID: 0000-0002-2576-7342
             \and
             In-Jae Yu \at
             Visual Display Business, Samsung Electronics Co., Ltd., Suwon, South Korea \\
             \email{injae.yu@samsung.com} \\
             ORCID: 0000-0001-9865-2194
             \and
             Heung-Kyu Lee \at
              School of Computing, Korea Advanced Institute of Science and Technology (KAIST), Daejeon, South Korea \\
                \email{heunglee@kaist.ac.kr}   
              \and
              Changick Kim \at
              School of Electrical Engineering, Korea Advanced Institute of Science and Technology (KAIST), Daejeon, South Korea \\
                \email{changick@kaist.ac.kr}   
  }
  
  \date{Received: 30 Aug 2021 / Accepted: 18 Apr 2022}

  \maketitle
  
  \begin{abstract}
  
  Detecting and localizing image manipulation are necessary to counter malicious use of image editing techniques.
  Accordingly, it is essential to distinguish between authentic and tampered regions by analyzing intrinsic statistics in an image.
  We focus on JPEG compression artifacts left during image acquisition and editing.
  We propose a convolutional neural network (CNN) that uses discrete cosine transform (DCT) coefficients, where compression artifacts remain, to localize image manipulation.
  Standard CNNs cannot learn the distribution of DCT coefficients because the convolution throws away the spatial coordinates, which are essential for DCT coefficients.
  We illustrate how to design and train a neural network that can learn the distribution of DCT coefficients.
  Furthermore, we introduce Compression Artifact Tracing Network (CAT-Net) that jointly uses image acquisition artifacts and compression artifacts.
  It significantly outperforms traditional and deep neural network-based methods in detecting and localizing tampered regions. 
  \end{abstract}
  
  \begin{figure*}[t]%
  \centering%
  \subfigure[Authentic]{%
    \includegraphics[width=0.13\linewidth]{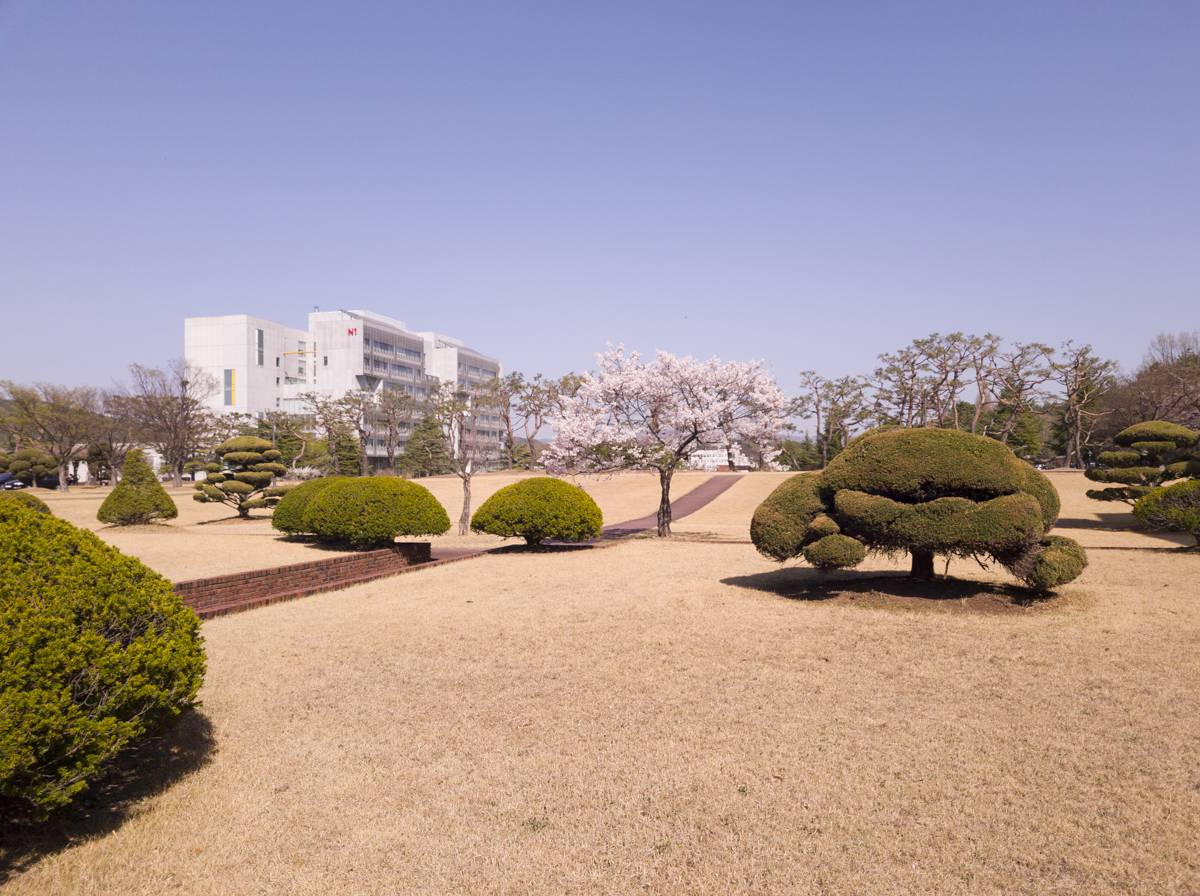}%
    \label{fig:cat_problem_a}
  }\hspace{-1mm}
  \subfigure[Forged]{%
    \includegraphics[width=0.13\linewidth]{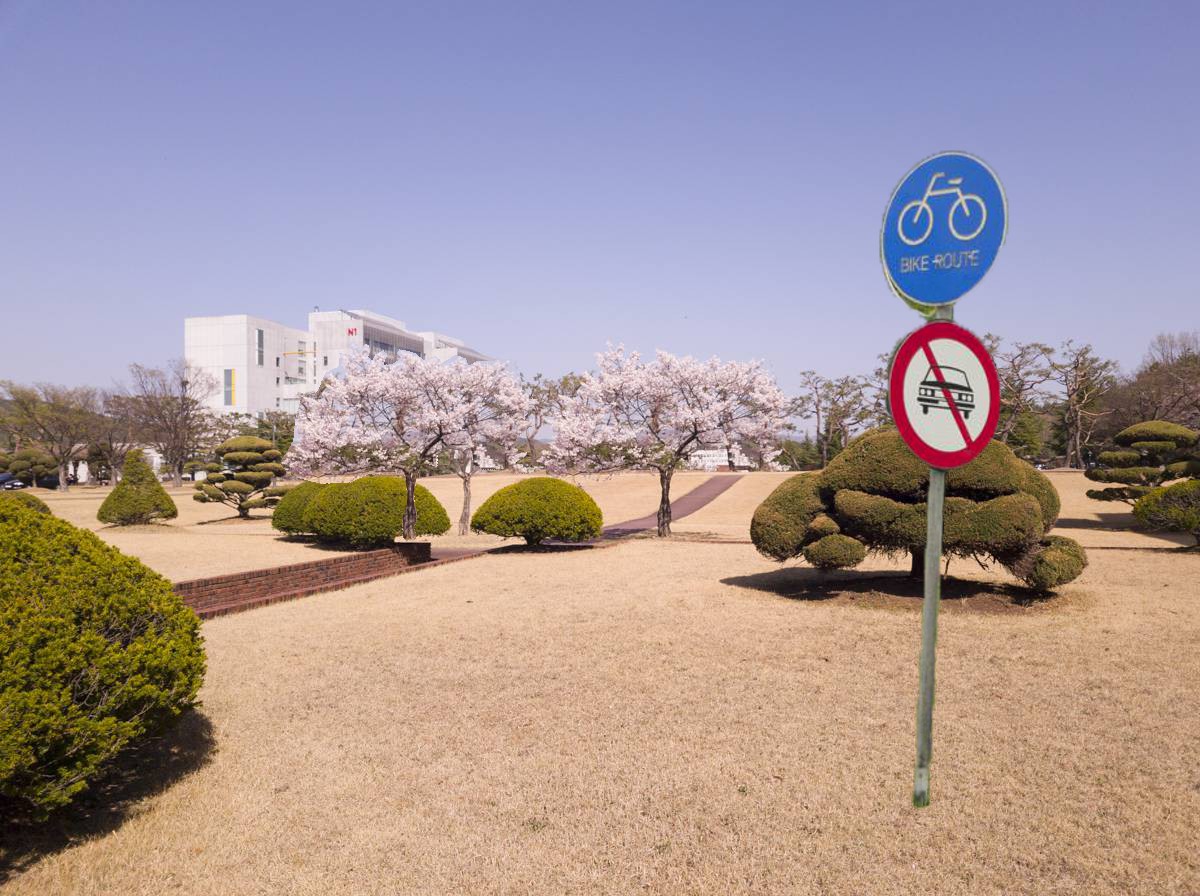}%
    \label{fig:cat_problem_b}
  }\hspace{-1mm}
  \subfigure[Ground truth]{%
    \includegraphics[width=0.13\linewidth]{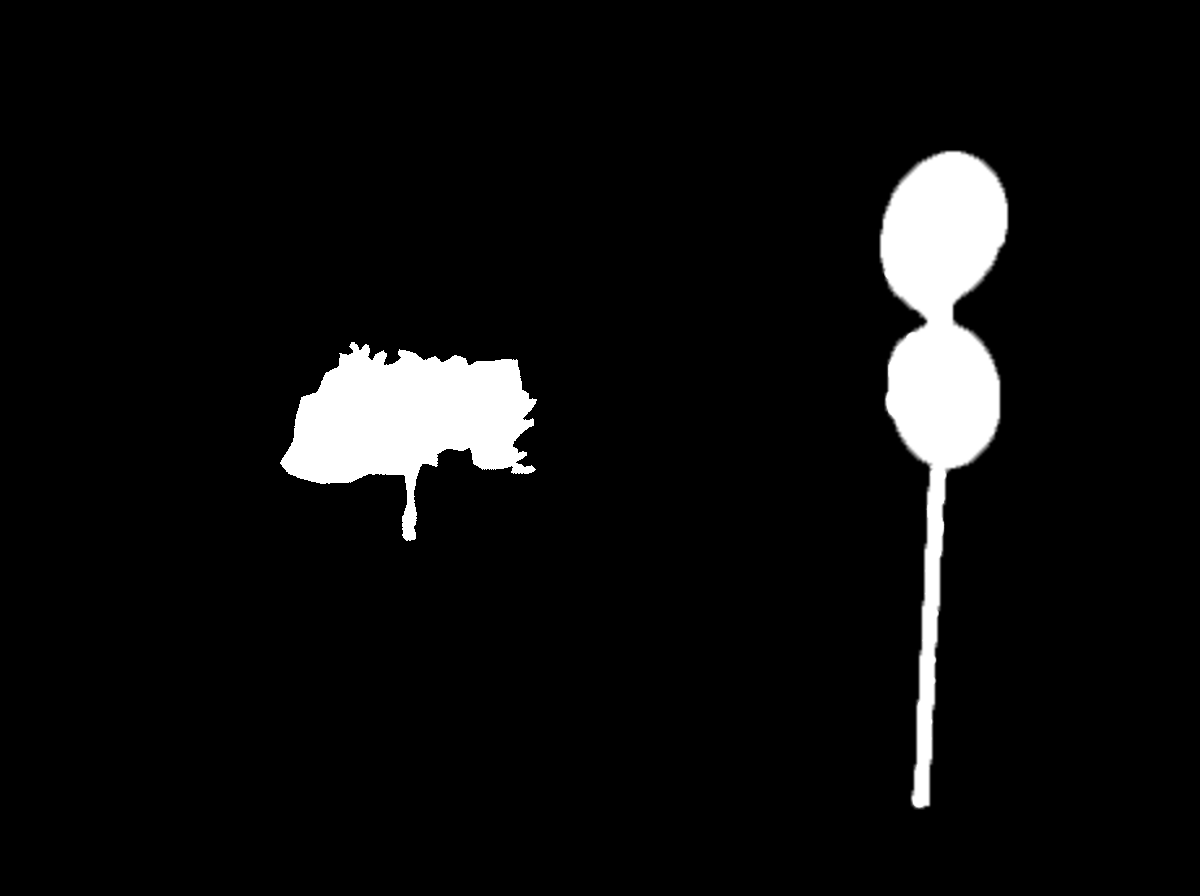}%
    \label{fig:cat_problem_c}
  }\hspace{-1mm}
  \subfigure[EXIF-SC]{%
    \includegraphics[width=0.13\linewidth]{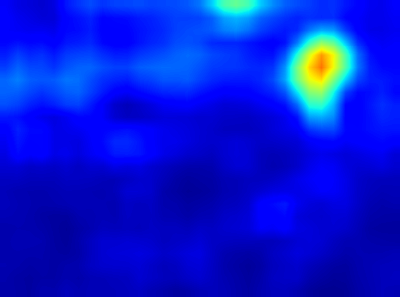}%
    \label{fig:cat_problem_d}
  }\hspace{-1mm}
  \subfigure[ManTra-Net]{%
    \includegraphics[width=0.13\linewidth]{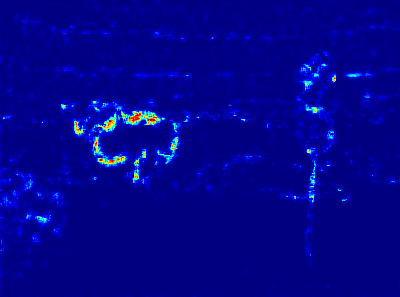}%
    \label{fig:cat_problem_e}
  }\hspace{-1mm}
  \subfigure[Noiseprint]{%
    \includegraphics[width=0.13\linewidth]{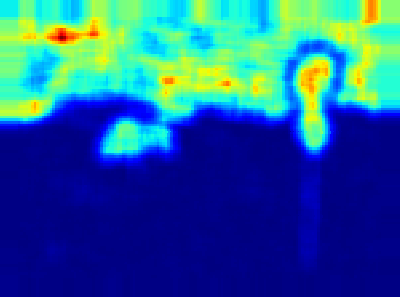}%
    \label{fig:cat_problem_f}
  }\hspace{-1mm}
  \subfigure[CAT-Net (Ours)]{%
    \includegraphics[width=0.13\linewidth]{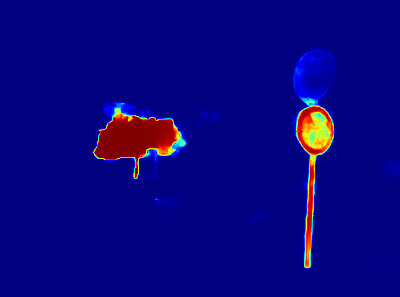}
    \includegraphics[width=0.021\linewidth]{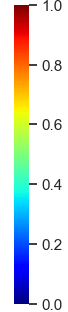}
    \label{fig:cat_problem_g}
  }
  
  \caption{Challenge of localizing manipulated regions from a JPEG image. Although many neural networks can trace noise precisely to detect manipulation, they are not ideal for capturing compression artifacts. The proposed approach considers RGB and DCT domains jointly to track visual clues and compression traces accurately. Given a possibly manipulated image (Fig.~\ref{fig:cat_problem_b}), this study predicts the manipulated region (Fig.~\ref{fig:cat_problem_g}). This study significantly outperforms state-of-the-art methods (Figs.~\ref{fig:cat_problem_d}, \ref{fig:cat_problem_e}, and \ref{fig:cat_problem_f}) in detecting and localizing forged regions.}
  \label{fig:cat_problem_formulation}%
  \end{figure*}
  
  \section{Introduction}
  \label{sec:cat_intro}
  
  With the advance of mobile devices and image editing software, image editing has become easy and popular. 
  Together with social networking services, edited images can be spread quickly.
  These changes enable people to create more beautiful selfies, reduce camera shake, place an unaccompanied friend in a group photo, remove undesired objects, and share these edited images with others.
  However, these advances cause social problems when edited images are used as false evidence or fake news. 
  An object-removed surveillance camera image might falsely confirm that a criminal was not at a crime scene or vice versa. 
  A fabricated photo suggesting a celebrity scandal might damage the celebrity's reputation.
  Therefore, to prevent malicious image manipulation, it is critical to detect them and localize forged regions.

  Among many image manipulation types, copy-and-pasting some regions onto an image either from the same image (\textbf{copy-move}) or another image (\textbf{splicing}) is one of the most popular and straightforward image editing techniques.
  Because these manipulations are applied to local regions, analyzing them is more challenging than kernel-based or pixel-level manipulation (\eg, hue modification, blurring, contrast enhancement, or brightness adjustment) applied to the global region.
  Furthermore, splicing and copy-move may not leave visual clues visible to the human eyes that consider the harmony between the pristine image and the objects to be pasted (Figs.~\ref{fig:cat_problem_a} and \ref{fig:cat_problem_b}).
  Consequently, in the last decade, many forensic approaches have been proposed to detect and localize image manipulation \citep{verdoliva2020media,korus2017digital}.

  A fundamental assumption underlying manipulated region detection and localization is that the image acquisition artifacts \citep{lukas2006digital} or JPEG compression artifacts \citep{wang_double_2016} of manipulated regions have different statistical properties from those of the pristine regions.
  An image acquired from a digital camera undergoes inherent internal processes. Thus, intrinsic statistical characteristics are left in digital images for each device and shooting setting.
  Moreover, most camera-equipped devices apply lossy compression (conventionally, JPEG) to the digital image for storage efficiency, leaving compression artifacts in the image.
  Characteristics of image acquisition and compression artifacts are consistently maintained within the media data if no manipulation occurs.
  Furthermore, the statistical characteristics of these artifacts can be changed when manipulation is applied.
  Image forensics aims to classify manipulated regions with different statistical fingerprints from pristine regions, so it is essential to understand detailed processes of image acquisition and JPEG compression.

  First, image acquisition artifacts refer to traces from the processes applied when creating a digital image from shooting a scene.
  The types of representative image acquisition artifacts are as follows: lens aberration \citep{yerushalmy2011digital}, sensor pattern noise \citep{lukas2006digital}, interpolation traces from the color filter array (CFA)~\citep{bammey2020adaptive}, and post-processing artifacts caused by color correction, white balance adjustment, and gamma correction \citep{swaminathan2008digital}.
  These artifacts are device- and setting-dependent fingerprints that accompany the image acquisition process and are difficult to distinguish with the human eye.
  In the field of multimedia forensics, rule-based, handcrafted feature-based, and data-driven approaches to capture changes in the statistical properties of each acquisition artifact have been studied~\citep{verdoliva2020media}. 
  These approaches are designed to detect and expose manipulated regions by revealing inconsistencies in fine acquisition artifacts.

  \begin{figure}[t]%
  \centering%
  \subfigure[Manipulated image]{%
   \includegraphics[width=0.48\linewidth]{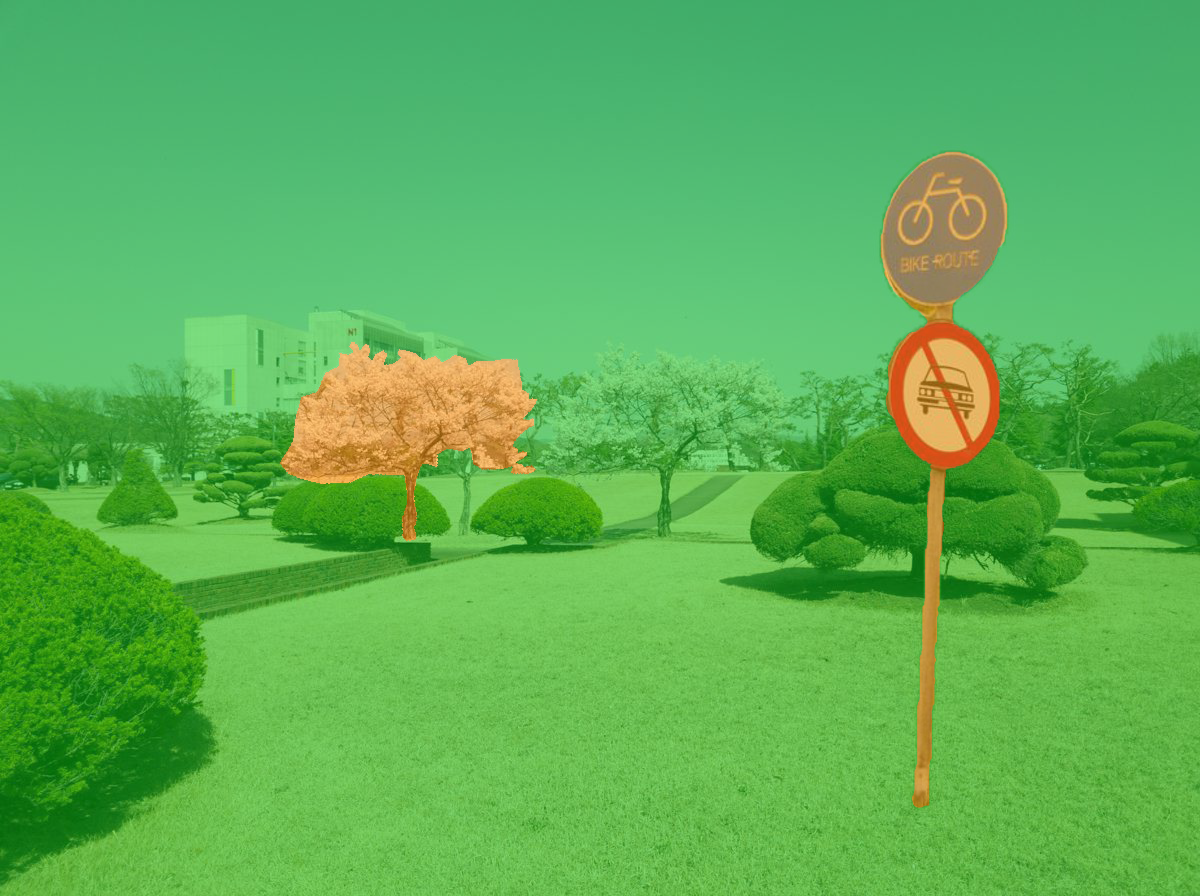}
   \label{fig:cat_histogram:a}
  }
  \subfigure[Authentic region]{%
    \includegraphics[width=0.48\linewidth]{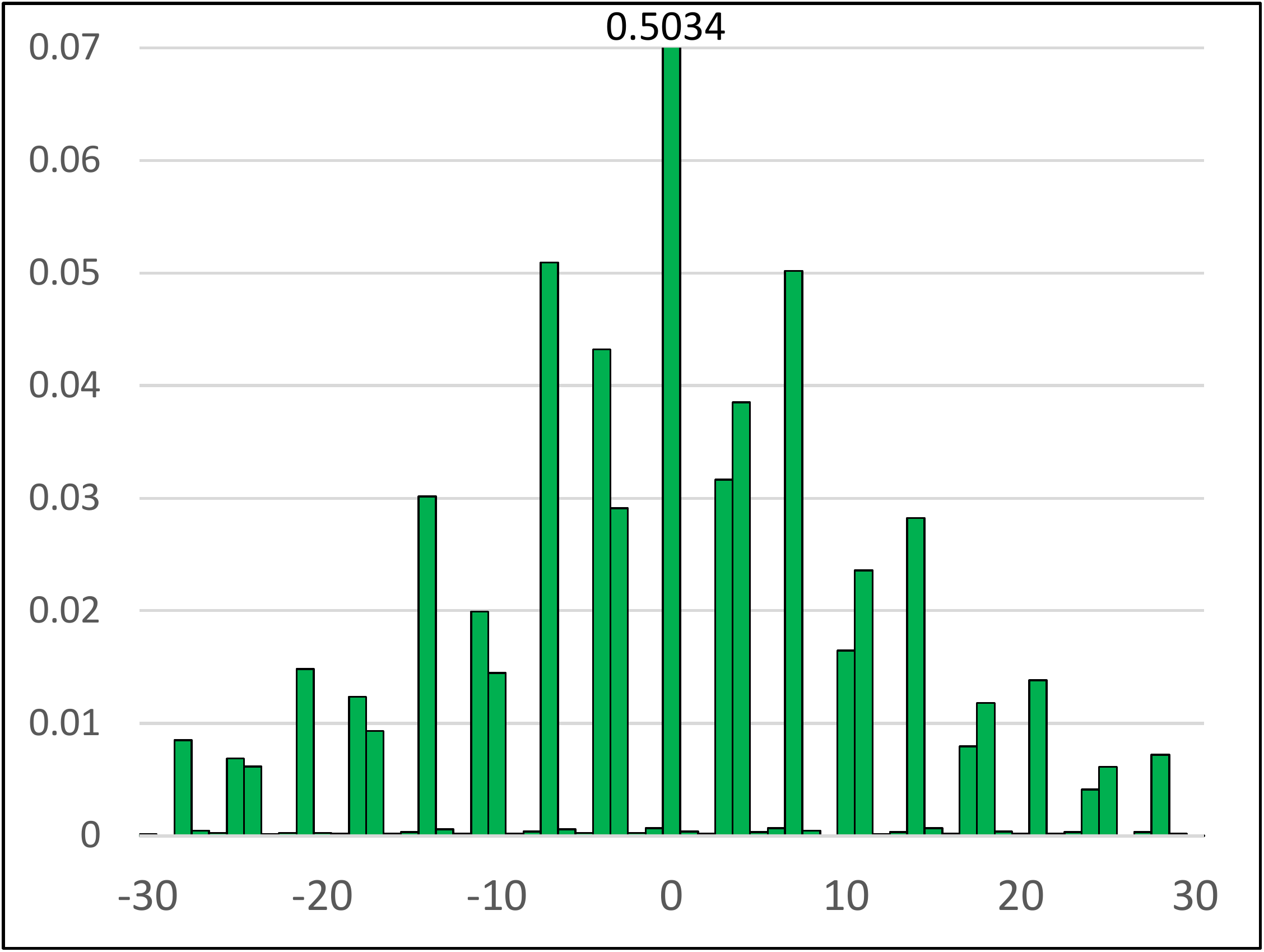}%
    \label{fig:cat_histogram:b}
  }
  \hspace{2mm}
  \subfigure[Manipulated region]{%
    \includegraphics[width=0.48\linewidth]{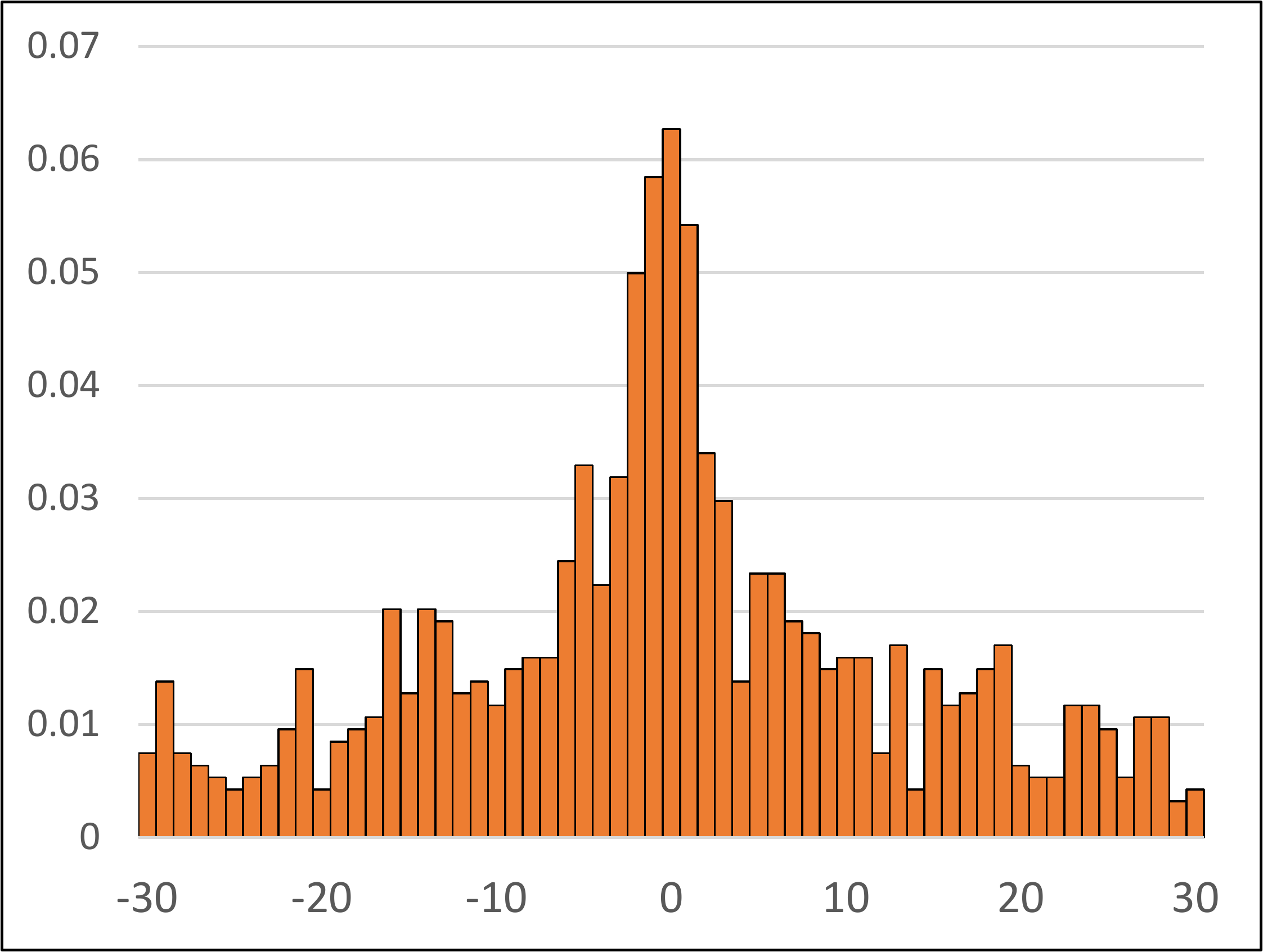}%
    \label{fig:cat_histogram:c}
  }
  \subfigure[Both regions]{%
    \includegraphics[width=0.48\linewidth]{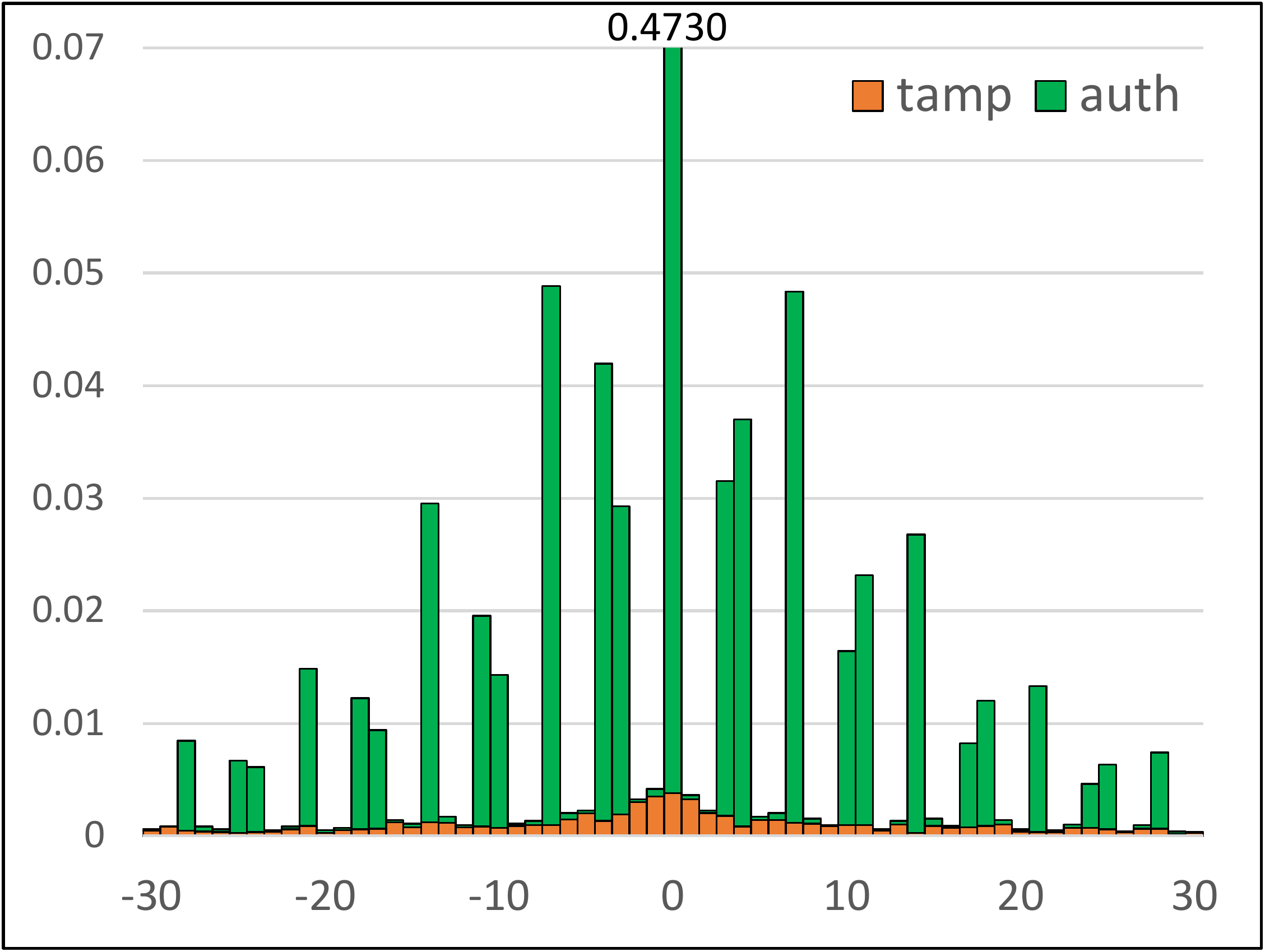}%
    \label{fig:cat_histogram:d}
  }
  \caption{Statistical differences between tampered and authentic regions. DCT histograms are obtained from Y-channel DCT coefficients at the frequency (1,1) for tampered and authentic regions separately. The x-axis is the DCT coefficient value and the y-axis is the relative frequency. The manipulated region follows a Laplacian distribution, whereas this distribution is interrupted for the authentic region. More in Sect.~\ref{sec:cat_related}.}
  \label{fig:cat_histogram}%
  \end{figure}

  Second, JPEG is the most actively used compression standard to reduce storage space, leaving subtle but distinct artifacts due to quantization-based compression applied to the discrete cosine transform (DCT) domain \citep{barni_aligned_2017}.
  In image forensic research, double JPEG detection, \ie, determining if a JPEG image has been compressed once or twice, is being actively studied \citep{wang_double_2016,park_double_2018,verma_block-level_2020}. This task helps localize the manipulation regions.
  A region pasted onto another image likely has a statistically different distribution of Y-channel DCT coefficients compared to the authentic region (Figs.~\ref{fig:cat_histogram:b} and \ref{fig:cat_histogram:c}).
  The authentic region is doubly compressed, first in a camera and again as part of the forgery, leaving periodic patterns in the histogram.
  The manipulated region follows a singly compressed distribution, based on the secondary quantization table (Sect.~\ref{sec:cat_rel_JPEG} and \cite{popescu_statistical_2004}).
  Therefore, the ability to explore these compression artifacts helps in inferring and localizing the manipulated region. 
  However, it is difficult to know in advance which region has been tampered with, \ie, what we observe is the sum of two histograms (Fig.~\ref{fig:cat_histogram:d}).

  Based on these observations of the two types of artifacts left in the manipulated image, we use both RGB and DCT domain information to detect and localize image manipulation. 
  We propose an end-to-end trainable neural network-based image manipulation detector named Compression Artifact Tracing Network (CAT-Net). It traces image acquisition artifacts and JPEG compression artifacts accurately.
  The RGB domain enables the network to explore and learn fine-grained visual artifacts such as sensor pattern noise, block artifacts, and other acquisition artifacts.
  The DCT domain is used to explore compression artifacts.
  
  However, supplying DCT coefficients directly to a convolutional neural network (CNN) is inadequate because the convolution throws away the spatial coordinates, which are crucial for DCT coefficients.
  Recently, \cite{yousfi_intriguing_2020} try to solve this problem using DCT volume representation in a steganalysis classification task. We adopt this representation in our network to learn the distribution of DCT coefficients. We demonstrate that this representation is also adequate for forgery localization tasks. Furthermore, the designed network includes only specially chosen network components to learn the image compression artifacts. Moreover, we propose a new pretraining method that uses double JPEG detection.

  This paper extends our previous study \citep{kwon2021cat}, which introduced a JPEG compression artifact-tracing method for image splicing detection. Whereas previous research only targeted splicing forgery, this paper also deals with copy-move forgery.
  New custom datasets are added to improve the performance further.
  More extensive experiments are performed with ten comparative methods, whereas only two methods were used in previous research. The results are reported with various metrics and newly added heatmaps. Finally, we released our code and trained weights publicly at \url{https://github.com/mjkwon2021/CAT-Net}.
  
  Our main contributions are summarized as follows: 
  \begin{itemize}[label={\tiny\raisebox{1ex}{\textbullet}}]
   \item We propose CAT-Net that learns compression artifacts based on DCT volume representation. This approach outperforms previous state-of-the-art networks using histogram representation in detecting double JPEG compression. Furthermore, we successfully transferred these weights to image manipulation detection and localization.
   \vspace{2mm}
   \item CAT-Net learns the distribution of DCT coefficients without losing spatial information to finely localize tampered regions. In contrast, previous histogram approaches lose spatial information and function only for classification. CAT-Net is the first neural network that accepts DCT coefficients directly into a segmentation network.
   \vspace{2mm}
   \item For the first time, CAT-Net localizes manipulated regions considering RGB and DCT domains jointly. The network captures image acquisition artifacts in the RGB domain and compression artifacts in the DCT domain. Extensive experiments with diverse benchmark datasets demonstrate that CAT-Net significantly outperforms state-of-the-art manipulation detectors.
  \end{itemize}

  The remainder of this paper is organized as follows.
  Section~\ref{sec:cat_related} explains forensic clues and reviews relevant previous studies.
  Section~\ref{sec:cat_proposed} proposes our forensic approach. Section~\ref{sec:cat_djpeg} explains a double JPEG pretraining scheme and evaluates CAT-Net in terms of learning compression artifacts.
  Section~\ref{sec:cat_imd} describes the main experiments, image manipulation detection and localization, and demonstrates the performance of CAT-Net. Section~\ref{sec:cat_conclusion} concludes the paper.

  \section{Related Work}
  \label{sec:cat_related}
  
  In this section, we review forensic clues including image acquisition artifacts and JPEG compression artifacts. We then introduce previous forensic approaches related to this study. Two types of inevitable artifacts remain in the digital image without manipulation: image acquisition artifacts and compression artifacts.
  These artifacts are essential forensic clues because their intrinsic properties differ before and after the manipulation process.
  
  \subsection{Image Acquisition Artifacts}
  \label{subsec:related_fc}
  
  \begin{figure}[t]
  \centering{\includegraphics[width=1.0\linewidth]{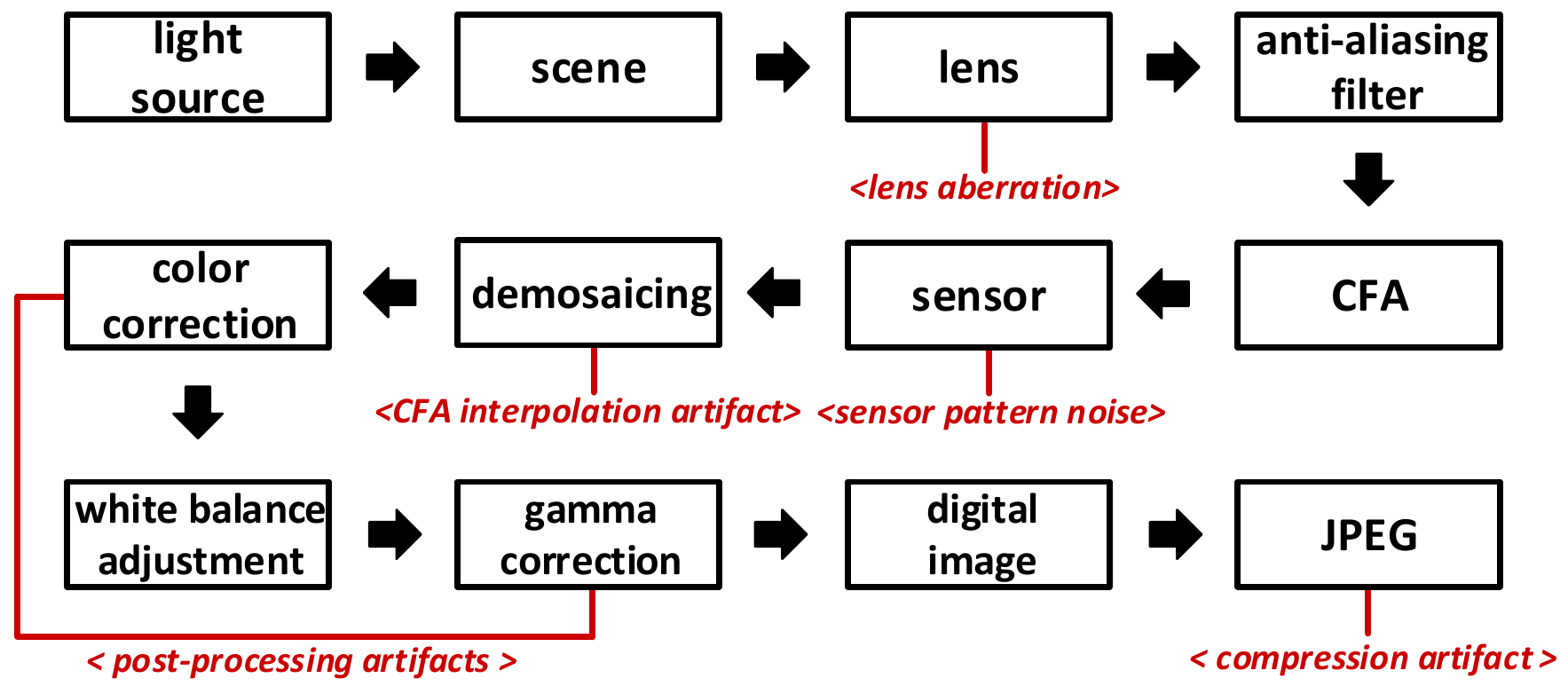}}
  \caption{Process of acquiring images from a digital camera. Red words illustrate artifacts exploited as forensic fingerprints.}
  \label{fig:cat_artifacts}
  \end{figure}

  Image acquisition artifacts denote fine artifacts generated when camera-equipped devices obtain a digital image. We can detect manipulated regions by distinguishing between authentic and tampered regions. Accordingly, we should partition the image into areas from the same image. Therefore, understanding camera-specific and capture-setting artifacts caused by the image acquisition process is advantageous.
  
  Figure~\ref{fig:cat_artifacts} illustrates the detailed processes of image acquisition from a digital camera.
  The terms in the arrow brackets refer to acquisition artifacts generated from a specific acquisition process.
  Before the light from the photographed scene reaches the sensor on the digital camera, it passes through the lenses, an anti-aliasing filter (\ie, optical low-pass filter), and a CFA.
  Because of the minor defects in the manufacturing process of a lens, the lens produces several types of image aberrations: spherical aberration, field curvature, lens radial distortion, and chromatic distortion.
  For source camera identification \citep{choi2006source} and forgery detection \citep{yerushalmy2011digital}, these lens aberrations can be used as forensic fingerprints.
  The light passed through the lens passes through an anti-aliasing filter, which reduces aliasing and moire patterns.
  Then, the light passes through a CFA before reaching the sensor. The CFA is a mosaic of color filters that block out a particular portion of the spectrum, inducing each pixel to detect only one specific color among red, green, and blue \citep{piva2013overview}.

  The sensor, composed of minimal addressable elements that collect photons and convert them into electrical signals, is a critical component of a digital camera.
  With an analog-to-digital converter on the imaging sensor, the voltages can be sampled to digital signals \citep{lukas2006digital}.
  The two sensor types include the charge-coupled device (CCD) and the complementary metal-oxide-semiconductor (CMOS), which both leave fine traces of sensor pattern noise \citep{piva2013overview}.
  The sensor pattern noise is caused primarily by imperfections during imaging sensor manufacturing. The main types of sensor pattern noise are fixed pattern and photo-response nonuniformity (PRNU).
  Because the pattern noise is an inherent property dependent on a specific camera model, it is actively used for image forensics as a distinct feature \citep{lukas2006digital,chierchia2014bayesian,korus2016multi}.

  For a CFA-based sensor (\eg, CCD or CMOS), the digitized sensor output is interpolated, exploiting the color interpolation process (\ie, demosaicing) to obtain the missing pixel values for the three-color layers \citep{piva2013overview}.
  In this process, CFA interpolation artifacts are applied to the image, and these traces can be used to detect image forgery \citep{bammey2020adaptive,choi2013estimation}.
  The output signal is then further processed based on post-processing, such as color correction, white balance adjustment, and gamma correction.
  These post-processing processes are fine corrections for perceptual quality, during which acquisition artifacts are added to the digital image.
  
  Finally, the digital image is written to the camera memory device in a user-selected image format. JPEG is a representative lossy compression technique for digital images that mitigates or removes the high-frequency components.
  This paper uses \textit{compression} to denote lossy compression (instead of loss-less compression, which is irrelevant for compression artifacts).
  The details of JPEG compression are introduced in the following subsection.

  \color{black}
  \subsection{JPEG Compression Artifacts}
  \label{sec:cat_rel_JPEG}
  In this subsection, we review the JPEG compression process and observe the double quantization artifacts left in the DCT domain. An input image is divided into non-overlapping $8 \times 8$ blocks, each block individually transformed using the DCT. In this paper, we consider only Y-channel DCT coefficients because chroma channels (\ie $C_b$ and $C_r$) are less useful for forensics. The DCT coefficients are then quantized using a single $8 \times 8$ quantization matrix.
  Quantization is an element-wise operation described as:
  \begin{equation}
      Q_{q_1}(u) = \left[{\frac{u}{{q_1}}}\right],
  \end{equation}
  where ${q_1}$ is the quantization step, $u$ is a value in the DCT domain, and $\left[\cdot\right]$ is a rounding operator. The quantized coefficients and the quantization table --- not the spatial domain pixels --- are saved in a JPEG file. The coefficients are dequantized when the image file is opened (\ie, during the JPEG decoding process):
  \begin{equation}
  \label{eq:dequantize}
      Q_{q_1}^{-1}(v) = {q_1}v,
  \end{equation}
  where $v$ is the quantized DCT coefficient.
  $Q_{q_1}$ is not mathematically invertible, so the loss of information occurs here.
  Double quantization can then be described as:
  \begin{equation}
      Q_{q_1,q_2}(u) = Q_{q_2} \left( Q_{q_1}^{-1} \left( Q_{q_1} \left(u\right) \right) \right) = \left[ \left[\frac{u}{q_1}\right] \frac{q_1}{q_2} \right],
  \end{equation}
  where $q_1$ is the primary quantization step and $q_2$ the secondary quantization step.
  
  We then investigate the relationship between an initial DCT coefficients histogram and a double compressed histogram. Assume a DCT coefficient in the $u_1$-th bin in the former is relocated in a $u_2$-th bin in the latter,~\ie, $Q_{q_1,q_2}(u_1)=u_2$.
  Then, the number of original histogram bins $n(u_2)$ contributing to bin $u_2$ in the double quantized histogram can be expressed as follows \citep{lin2009fast}:
  \begin{equation}
  \label{eq_hist_bin}
      n(u_2) = q_1 \left( \floor*{\frac{q_2}{q_1} \left( u_2 + \frac{1}{2}\right)} - \ceil*{\frac{q_2}{q_1} \left( u_2 - \frac{1}{2}\right) } + 1 \right),
  \end{equation}
  where $\floor{\cdot}$ is the flooring operator and $\ceil{\cdot}$ is the ceiling operator.
  Based on Eq. (\ref{eq_hist_bin}), $n(u)$ is periodic with period $q_1 / \text{gcd}(q_1,q_2)$ where $\text{gcd}$ is the greatest common divisor. Therefore, the double compressed region has periodic patterns in the histograms of quantized DCT coefficients. 
  For example, Fig.~\ref{fig:cat_histogram} illustrates a double compressed image with quality factor 70 followed by 90 and the effects of double quantization at frequency (1, 1). 
  Then, $q_1=T_{70}(1,1)=7$ and $q_2=T_{90}(1,1)=2$ where $T_{x}(i,j)$ is the value of the $(i,j)$ component of the quantization table with the quality factor $x$, where $i,j=0,...,7$. Thus, based on the Eq. (\ref{eq_hist_bin}), 
      $
          n(7k)=7,
          n(7k+1)=0,
          n(7k+2)=0,
          n(7k+3)=7,
          n(7k+4)=7,
          n(7k+5)=0, \text{ and }
          n(7k+6)=0
      $
  where $k$ is an integer.
  This values coincide with the observation that specific bins are empty in Fig.~\ref{fig:cat_histogram:b}. This periodic pattern is an example of many double compression effects (More in Sect.~\ref{subsec:cat_related_for_comp}). 
  
  The above reasoning assumed that quantization uses rounding to the nearest integer with tie-breaking toward positive infinity. However, different operations such as rounding toward zero can be used depending on camera manufacturers or image editing software \citep{agarwal2018jpeg, butora2020steganography}. 
  Furthermore, information loss occurs during decoding by rounding after inverse DCT is applied and truncating to the proper image pixel range [0, 255].
  The precision of the DCT transform also impacts the distribution of coefficients \citep{lukas_estimation_2003}. 
  Accordingly, quantization artifacts in real-world implementations are diverse and should be handled with care.

  The manipulated and authentic portions exhibit different statistical distributions in the DCT histogram. The authentic regions are compressed twice. The tampered region is treated as single compression because the $8 \times 8$ grid used in the second compression is likely misaligned with the primary compression grid (with probability $\frac{63}{64}$). Even when the two grids align, blocks containing the boundary of the pasted object have both authentic and tampered pixels, so these blocks do not follow the double compression rules~\citep{wang_double_2016}.

  \subsection{Image Forensics Using Image Acquisition Artifacts}
  \begin{table}[t]
  \caption{Summary of image manipulation detection and localization methods. Top: methods not using deep learning, bottom: methods using deep learning.}
  \centering
  \resizebox{\linewidth}{!}
  {
      \begin{tabular}{llll}\\
      \hline
      Method & Final Decision & Forensic Clue & Localization\\
      \hline
      \cite{lukas_estimation_2003} & 2-layer neural network & DCT histogram & Image-level \\
      \cite{ye2007detecting} & Rule-based algorithm & DCT histogram & Block-level \\
      \cite{fu2007generalized} & SVM & First digit distribution of DCT coef. & Image-level  \\
      \cite{lin2009fast} & SVM & Double compression artifacts & Block-level \\
      \cite{mahdian2009using} & Block merging & Noise inconsistency & Block-level \\
      \cite{amerini2011sift} & Clustering & SIFT descriptor & Object-level \\
      \cite{bianchi2012image} & Mathematical modeling & Non-aligned requantization artifact & Block-level\\
      \cite{ferrara2012image} & Mathematical modeling & Demosaicing artifact & Block-level \\
      \cite{lyu2014exposing} & Mathematical modeling & Noise inconsistency & Block-level \\
      \cite{iakovidou2018content} & Rule-based algorithm & JPEG grid inconsistency & Block-level\\
      \cite{nikoukhah2019jpeg} & Rule-based algorithm & Number of zeros in the $8\times8$ DCT blocks & Block-level \\
      \hline
      & & & \\
      \hline
      Method & Backbone Network & Forensic Clue & Localization\\
      \hline
      \cite{wang_double_2016} & CNN & DCT histogram & Image-level \\
      \cite{barni_aligned_2017} & CNN & DCT histogram & Image-level \\
      \cite{park_double_2018} & CNN & DCT histogram + quantization table & Image-level \\
      \cite{bayar2018constrained} & CNN & Noise residual with constrained layer & Image-level  \\
      \cite{zhou_learning_2018} & Faster R-CNN & Visual tampering artifact+Noise & Object-level  \\
      \cite{boroumand_deep_2019} & CNN & Noise residual with unpooled layer & Image-level  \\
      \cite{huh_fighting_2018} & SiameseNet & EXIF metadata inconsistency & Block-level \\    
      \cite{bi2019rru} & U-Net & Image essence property & Pixel-level \\
      \cite{wu_mantra-net_2019} & VGG+ConvLSTM & Anomalous feature & Pixel-level \\
      \cite{kniaz_point_2019} & GAN & Semantic inconsistency & Pixel-level \\
      \cite{cozzolino2019noiseprint} & SiameseNet & Camera model fingerprint & Pixel-level \\
      \cite{bammey2020adaptive} & CNN & Local CFA inconsistency & Block-level  \\
      \cite{marra2020full} & Xception & Spatial anomalies with noise residual  & Image-level \\
      \cite{hu2020span} & VGG & Anomalous feature & Pixel-level  \\
      \cite{liu2020exposing} & DenseNet & Noise and JPEG discrepancies & Image-level \\
      Ours & HRNet & Acquisition and compression artifacts & Pixel-level \\
      \hline
      \end{tabular}
  }
  \label{tab:forensics_approaches}
  \end{table}

  
  Image forensics aims to verify the authenticity of media content by detecting and exploring manipulation artifacts.
  It is challenging to localize and detect manipulation applied to local regions (\eg, splicing or copy-move) and related studies are steadily progressing. Table~\ref{tab:forensics_approaches} summarizes historic image forensic approaches.
  The forensic clues used by each approach are categorized primarily into image acquisition and JPEG compression artifacts. This subsection reviews previous studies that explore traces of image acquisition.
  These studies either use acquisition artifacts directly as forensic features or explore low-level features to detect statistical changes on acquisition artifacts caused by image manipulation.

  \cite{mahdian2009using} propose a forgery localization method using local noise standard deviation estimated based on tiling the high-pass wavelet coefficient.
  \cite{amerini2011sift} use a scale-invariant feature transform (SIFT) to detect copy-move forgery; the pairs of SIFT descriptors between the pristine and manipulated regions are selected using a clustering algorithm.
  \cite{ferrara2012image} perform block-based forgery detection exploring demosaicing artifacts, a subtle deformation applied to the original CFA pattern during the forgery process.
  \cite{lyu2014exposing} formulate blind noise estimation as an optimization problem and detect local noise inconsistency to localize region forgery.

  Inspired by computer vision tasks that have achieved significant progress after adopting CNNs, CNNs are actively exploited in image forensics to localize forged areas and detect fine-grained manipulation clues~\citep{verdoliva2020media, nam2020deep, yu2020manipulation}.
  \cite{bayar2018constrained} propose a constrained layer-based network that jointly suppresses the content of a given image and adaptively learns features from noise-like signals generated by image manipulation.
  \cite{zhou_learning_2018} place SRM kernel \citep{fridrich2012rich} as a pre-processing layer and uses the Faster R-CNN \citep{ren2015faster} architecture to detect the manipulated area in units of object-level.
  \cite{boroumand_deep_2019} place unpooled layers at the early part of the network, which extracts rich features of low-level signals and illustrates excellent performance for steganalysis.
  \cite{huh_fighting_2018} propose a self-supervised approach to train a model and explores the inconsistency of EXIF metadata.
  Their research exhibits outstanding performance in localizing manipulation but requires significant computation to compute the consistency for every patch pair.
  \cite{bi2019rru} frame manipulation localization as a segmentation problem. They design a neural network based on U-Net \citep{ronneberger2015u} and analyze the conventional semantic property by providing an RGB pixel image as input to the network.

  \cite{wu_mantra-net_2019} design a ManTra-Net that extracts features using the SRM kernel and constrained layer for preprocessing and performs pixel-wise anomaly detection.
  Their research classifies various types of manipulation successfully. However, the performance of the forgery localization is not robust to JPEG compression because it uses compression as one of the manipulation types.
  \cite{kniaz_point_2019} propose a generative adversarial network-based framework for training a discriminative segmentation model to localize manipulated regions.
  \cite{cozzolino2019noiseprint} propose an approach to extract intrinsic noise of a camera model (\ie, Noiseprint), where the content of a given image is suppressed and acquisition artifacts are enhanced.
  Their study explores anomalies with respect to the dominant pristine model for localizing manipulated parts.

  \cite{bammey2020adaptive} exploit an unsupervised CNN that learns to explore the underlying pattern of CFA interpolation artifacts and detects suspicious regions by identifying local mosaic inconsistencies.
  \cite{marra2020full} present a framework comprising of three phases --- patch-wise feature extraction, image-wise feature aggregation, and global decision --- that enables the use of rich features gathered at full resolution from the whole image.
  \cite{hu2020span} present a local self-attention block-based CNN that models and establishes the spatial relationship between patches at multiple scales to capture forensic fingerprints in forgery localization.
  \cite{liu2020exposing} propose a fusion network that concentrates on learning low-level features and explores forensic hypotheses such as noise and JPEG discrepancies.

  \subsection{Image Forensics Using JPEG Compression Artifacts}
  \label{subsec:cat_related_for_comp}
  
  %
  
  In this subsection, we review previous forensic methods exploiting forensic clues caused by JPEG compression.
  \cite{lam2000mathematical} mathematically illustrate that histograms of JPEG DCT coefficients follow the Laplacian distribution. 
  Image editing conventionally involves additional compression and breaks the distribution, leaving compression traces in the image. 
  Therefore, JPEG compression artifacts have been used as important fingerprints in image forensics.
  
  \cite{lukas_estimation_2003} observe a fundamental characteristic left in the DCT domain when an image is forged.
  A pasted portion of a forged image likely exhibits traces of single compression, while the rest of the authentic region exhibits signs of double compression.
  The researchers present properties of missing values and double peaks in a histogram of DCT coefficients to detect double compression and estimate the primary quantization table. 
  \cite{fu2007generalized} observe that the distribution of the first digits of the DCT coefficients follows Benford's law, which is violated if the image is double compressed. This law is used for Q-factor estimation and double JPEG detection.
  \cite{ye2007detecting} use inconsistency of JPEG blocking artifacts to detect forgeries.
  A blocking artifact measure is calculated based on the estimated quantization table using the power spectrum of the DCT coefficient histogram.
  \cite{lin2009fast} use JPEG double quantization effects such as periodic peaks and valleys in the DCT histogram to detect image forgeries automatically at the scale of $8 \times 8$ blocks.
  A block posterior probability map computed from histograms is thresholded to differentiate the tampered and authentic regions.
  \cite{bianchi2012image} design a unified statistical model characterizing the DCT coefficients for both aligned and nonaligned double JPEG compression.
  The model is used to compute a likelihood map indicating the probability of each DCT block being doubly compressed.
  \cite{iakovidou2018content} use JPEG grid alignment abnormalities for forgery detection.
  The method evaluates multiple grid positions using a fitting function, where lower contribution areas are identified as grid discontinuities.
  \cite{nikoukhah2019jpeg} perform global grid detection via determining the likeliest JPEG blocks containing the largest number of zero coefﬁcients to detect grid non-alignment caused by splicing.

  In the deep learning era, there have been subsequent studies on double JPEG detection using DCT histograms to generate neural network features.
  \cite{wang_double_2016} are the first to use histogram features as input to a CNN for double JPEG detection.
  They also achieve forgery localization by integrating image-level classification results using an overlapping stride of 8 pixels.
  \cite{barni_aligned_2017} integrate histogram computation as part of a CNN, allowing GPU to construct histogram features in parallel.
  They also improve the structure of the CNN using two-dimensional (2D) convolutions instead of one-dimensional (1D) convolutions.
  \cite{park_double_2018} improve classification performance by appending a reshaped quantization table in fully connected layers.

  \begin{figure*}[t]
  \centering{\includegraphics[width=1.0\linewidth]{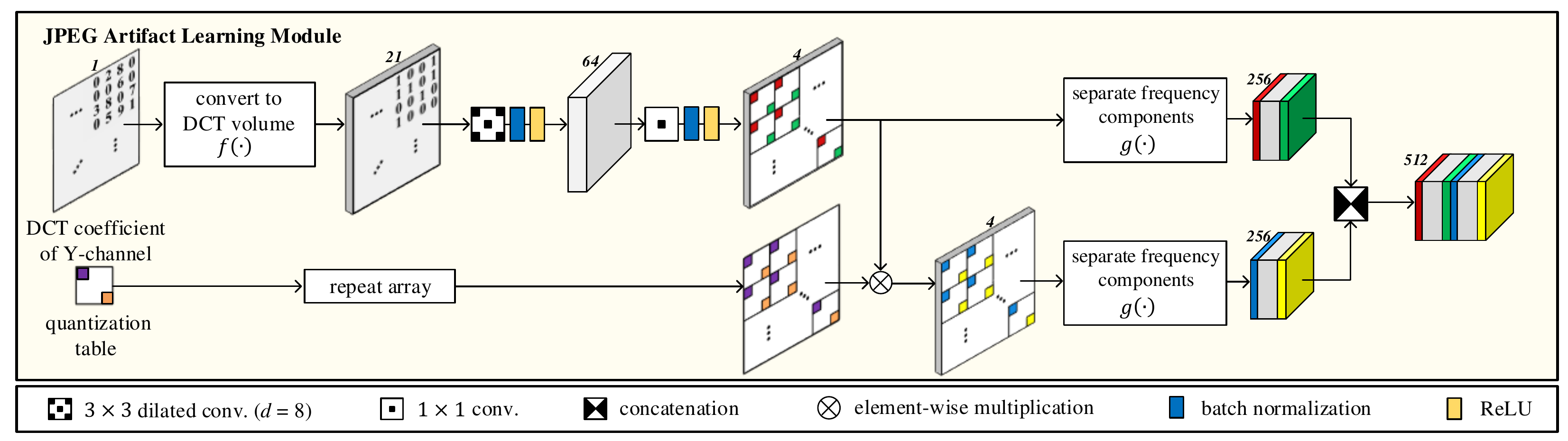}}
  \caption{Proposed JPEG artifact learning module architecture. The DCT volume conversion $f(\cdot)$ and the frequency component separation $g(\cdot)$ are depicted in Eqs.~\ref{eq:binary_volume} and~\ref{eq:cat_compoment}, respectively.}
  \label{fig:cat_JPEG}
  \end{figure*}

  CNN-based approaches with a DCT histogram are confined to image-level classification, primarily because using a DCT histogram requires a fixed size input and removes spatial information for localization.
  Previously, using a DCT histogram was mandatory because CNNs could not learn from naive DCT coefficients due to their predominantly decorrelated and locally heterogeneous nature.
  This study is the first to use a segmentation model based on DCT coefficients, which is possible due to the DCT volume replacing the DCT histogram and carefully designed network components. Moreover, we pretrain our network in a double JPEG detection task to produce rich initialization for learning compression artifacts left in the DCT coefficients.

  \section{Proposed Method}
  \label{sec:cat_proposed}
  
  We describe how to extract features over DCT coefficients to learn their distributions using standard CNN components. Accordingly, we propose a JPEG artifact learning module (Fig.~\ref{fig:cat_JPEG}) that can be placed at the starting point of a CNN. Furthermore, we propose CAT-Net, a complete end-to-end image manipulation detection network.
  CAT-Net comprises an RGB stream, DCT stream, and fusion stage (Figs.~\ref{fig:cat_architecture} and \ref{fig:cat_basicblock_fusion}). It accepts RGB pixels, DCT coefficients, and a quantization table as network inputs and outputs a probability map of each pixel being tampered with.
  We first describe four key points that enable a CNN to learn the distribution of DCT coefficients: DCT volume representation, frequency-wise operations, grid-aligned cropping, and transfer learning from double JPEG detection.
  We then describe the detailed network architecture.
  Finally, we describe how CAT-Net processes non-JPEG images.

  \subsection{DCT Volume Representation}
  As explained in Sect.~\ref{sec:cat_intro}, CNNs cannot automatically learn the compression artifacts from raw DCT coefficients because the convolution assumes a translation-invariant property and handles every coefficient the same. However, the spatial coordinates are critical for DCT coefficients. Thus, we convert the input array of DCT coefficients, $\mathbf{M}$, to a binary volume~\citep{yousfi_intriguing_2020} using a transformation $f:\mathbb{Z}^{H \times W} \to {\{0,1\}}^{(T+1) \times H \times W} $ such that
   \begin{equation}
   \label{eq:binary_volume}
   f(\mathbf{M})_{t,i,j}=
   \begin{cases}
      1, & \text{if }\, abs(clip(\mathbf{M}))_{i,j}=t\\
      0, & \text{otherwise}
  \end{cases},\end{equation}where $clip(\cdot)$ clips the array element-wise into the interval $[-T, T]$ and $abs(\cdot)$ takes element-wise absolute values.
  The DCT coefficients are recorded in channel indices with 0 or 1.
  
  The $clip(\cdot)$ function is due to memory constraints. A larger $T$ enables capturing a broader range of histogram bins (Fig.~\ref{fig:cat_histogram}) but requires more GPU memory. We chose $T$ to be 20, experimentally. 
  The $abs(\cdot)$ function is due to the symmetry of the DCT histogram as depicted in Eq.~\ref{eq_hist_bin}. With the identity $\floor{-x}=-\ceil{x}$, we obtain $n(-u)=n(u)$. Thus, information loss caused by taking absolute values is negligible, but the feature map size becomes almost half.
  
  For manipulation localization in JPEG images, the DCT volume representation is more accurate than the DCT histogram, which detects double JPEG compression~\citep{wang_double_2016, barni_aligned_2017, park_double_2018}.
  Whereas the DCT histogram merges information patch-wise and loses its visual representation, the DCT volume maintains image resolution suitable for prediction at the pixel level.
  Nevertheless, the ability to extract statistical information is an improvement over DCT histograms (Sect.~\ref{sec:cat_djpeg}).
  
  The DCT histogram is the result of applying global average pooling to the DCT volume. Thus, the DCT volume is a feature before losing location information.
  Furthermore, the convolution on this representation produces much richer statistical features such as co-occurrence. For example, consider a $3 \times 3$ kernel $K\in\mathbb{R}^{(T+1)\times3\times3}$, where all the elements in $K$ are zero except $K[m,1,1]=1$, $K[n,1,2]=1$, and $m,n\in\{0,1,...,T\}$. If a convolution using the kernel $K$ is applied to the DCT volume and global average pooling is followed, the horizontal co-occurrence for coefficient pairs $(m,n)$ is computed.
  In contrast to JPEG image steganalysis, which evaluates the probability of whole image manipulation, our goal is localizing manipulation.
  Therefore, we aim to extract features among different DCT blocks and use convolutions with a dilation of 8, which enables frequency-wise operations.

  \subsection{Frequency-wise Operations}
  \label{subsec:cat_freq_op}
  In contrast to RGB pixels, DCT coefficients represent different frequencies depending on where they are located. The DCT coefficient at $(x,y)$ represents frequency $(x\text{ mod }8,y\text{ mod }8)$ of the $\left( \floor{\frac{x}{8}}, \floor{\frac{y}{8}} \right)$ image subblock.
  Conventional convolutions (with stride one) mix these frequency components. All operations should be performed on a frequency-wise basis to avoid this. Namely, these include an $8 \times 8$ convolution with a dilation of 8, a $1 \times 1$ convolution, quantization table multiplication, and frequency component separation (Fig.~\ref{fig:cat_JPEG}).
  The $8 \times 8$ convolution with a dilation of 8 operates on the same frequencies because DCT coefficients consist of $8 \times 8$ blocks. The $1 \times 1$ convolution is also valid because it does not mix frequency components.
  A quantization table is used to help the network learn compression history. 
  
  \cite{park_double_2018} are the first to use a quantization table in fully connected layers. However, because our network is fully convolutional, we cannot follow their approach. We solve this problem by mimicking the JPEG decoding process. The quantization table is multiplied element-wise with a feature map. This approach uses the role of quantization tables for dequantizing quantized coefficients (Eq.~\ref{eq:dequantize}). Quantized and dequantized feature maps are both used in our module.
  
  Frequency component separation $g:\mathbb{R}^{C \times H \times W} \to \mathbb{R}^{64C \times \floor{\frac{H}{8}} \times \floor{\frac{H}{8}}} $ is an index changing mapping that can be implemented using only reshaping and permutation:
  \begin{equation}
  \label{eq:cat_compoment}
    g(\mathbf{M})_{c,i,j} = \mathbf{M}_{\floor*{\frac{c}{64}},8i+\floor*{\frac{c\text{ mod }64}{8}},8j+(c\text{ mod }8)},
  \end{equation}where $c,i,j$ starts from 0.
  After frequency component separation, the feature maps can be used without special care,~\ie, conventional $3\times 3$ convolutions may follow.

  \begin{figure*}[t]
  \centering{\includegraphics[width=1.0\linewidth]{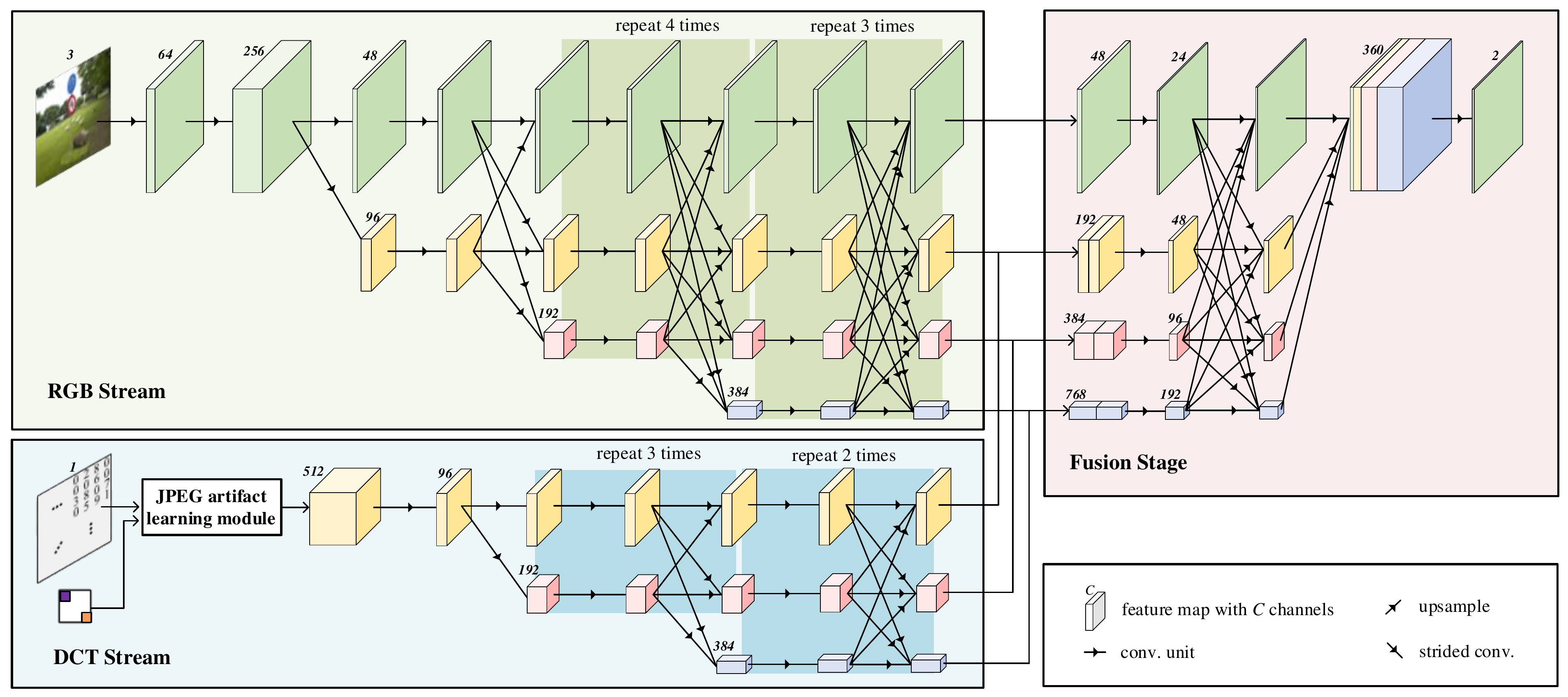}}
  \caption{Proposed CAT-Net architecture, including an RGB stream, a DCT stream, and a final fusion stage. The RGB stream takes RGB pixels and the DCT stream takes Y-channel DCT coefficients and a Y-channel quantization table as inputs. JPEG artifact learning module is depicted in Fig.~\ref{fig:cat_JPEG}.}
  \label{fig:cat_architecture}
  \end{figure*}
  
  \begin{figure*}[t]%
  \centering%
  \subfigure[Basic block]{%
    \includegraphics[height=1.1in]{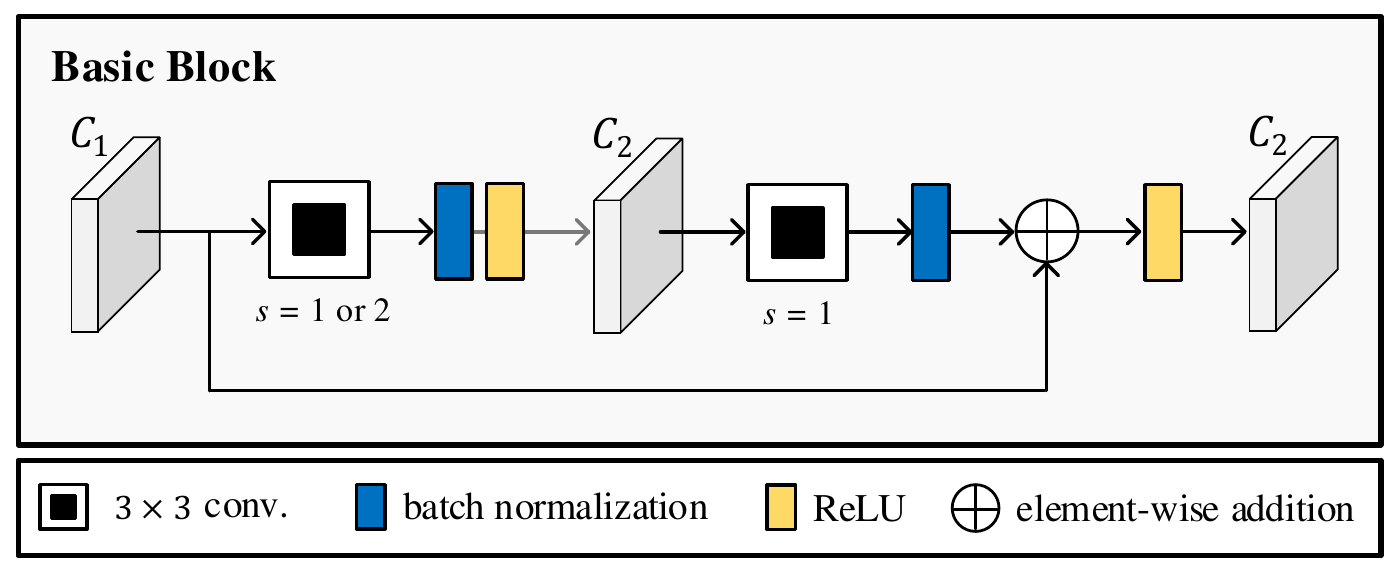}%
    \label{fig:cat_basicblock}
  }\hspace{-1mm}
  \subfigure[Fusion unit]{%
    \includegraphics[height=1.1in]{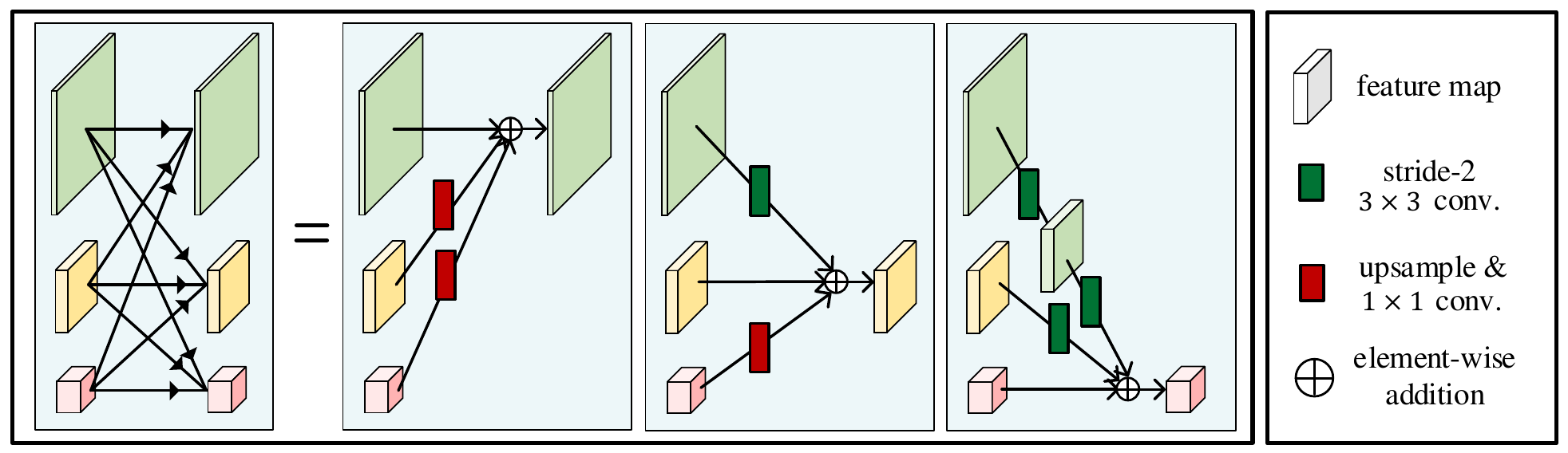}%
    \label{fig:cat_fusion_unit}
  }
  \caption{Elements in the proposed network. A convolutional unit in Fig.~\ref{fig:cat_architecture} mostly consists of four consecutive basic blocks. Fusion unit fuses multi-resolution feature maps by summing them after matching resolutions.}
  \label{fig:cat_basicblock_fusion}%
  \end{figure*}

  \subsection{Grid-aligned Cropping}
  \label{sucsec:cat_grid}
  Deep neural networks take input images of the same fixed size when training to construct a batch. Conventional computer vision networks use resizing or random cropping to satisfy this constraint. 
  With DCT coefficients, these two methods cannot be used because they destroy position information. 
  We propose a new cropping method that can be used for DCT coefficients. 
  Images should be cropped in a grid-aligned scheme to enable these components to function correctly. Given an input image $\mathbf{M}$ and crop size $h \times w$, the conventional cropped image $h(\mathbf{M})$ can be represented as:
  \begin{equation}
  h(\mathbf{M})=\mathbf{M}[i:i+h,j:j+w],
  \end{equation}
  where NumPy index slicing is used.
  Grid-aligned cropping requires $h,w,i,j$ to be the multiples of eight. This simple remedy enables the neural network components described in Sect.~\ref{subsec:cat_freq_op} to function.
  With grid-aligned cropping, each feature map channel after frequency component separation represents one frequency. For example, the first channel corresponds to frequency (0, 0), the second channel corresponds to frequency (0, 1), and so on. If conventional random cropping were used instead, frequency components would be distributed over all channels and the subsequent convolution could not distinguish frequencies.

  \subsection{Pretraining on Double JPEG Detection}
  It is common practice in the deep learning literature to start training using pretrained weights from a similar task. People often use pretrained weights from image classification for semantic segmentation, especially from ImageNet \citep{krizhevsky_imagenet_2012}. We also initialize the RGB stream pretrained on ImageNet to extract visual clues more efficiently. However, being the first study to use DCT coefficients as input to a segmentation network, there is no ``common practice'' to pretrain the DCT stream. We introduce a new pretraining scheme on a double JPEG detection task, classifying single and double compressed JPEG images. 
  
  The DCT stream is pretrained with various quantization tables on a double JPEG detection task to learn to handle real-world compression artifacts. 
  The pretrained weights are transferred to the image manipulation detection task. Obtaining datasets with various compression parameters for double JPEG detection is much easier than obtaining forgery images with ground truth masks. Ablation studies demonstrate that this pretraining scheme helps the network train faster and achieve higher detection performance (Sect.~\ref{subsec:cat_abl}).

  \subsection{Network Architecture}
  This subsection describes how we design the CAT-Net structure.
  CAT-Net consists of the RGB stream, DCT stream, and fusion stage. The RGB stream takes an RGB image as input and learns image acquisition artifacts, such as sensor pattern noise, EXIF metadata, blocking artifacts, or visual content itself.
  The DCT stream takes raw DCT coefficients and a quantization table obtained from the JPEG header as inputs and learns compression artifacts. 
  The network is built on top of HRNet~\citep{wang_deep_2020}. The structure of the RGB stream is HRNet itself. The DCT stream is a three-resolution variant of HRNet, replacing the first stage with our JPEG artifact learning module. 
  
  We adopt HRNet in a forensic task for the first time because it maintains high-resolution representations through the entire process, enabling us to capture the overall picture without losing fine artifacts essential for forensic investigations. 
  With HRNet as the backbone, CAT-Net can acquire fine-grained forensic clues and learn the correlation among different regions.
  In addition, the HRNet feature map sizes are well suited to tracing JPEG artifacts. 
  Because DCT is applied in $8 \times 8$ blocks, the minimum resolution the DCT stream can predict is $8 \times 8$, which is $\frac{1}{8}$th that of the input size (depicted in the yellow feature map in Fig.~\ref{fig:cat_architecture}). This resolution can be easily joined by concatenation with the second resolution in the RGB stream. 
  Furthermore, HRNet uses stride-2 convolution to downsample feature maps and does not use pooling layers. Recent studies have demonstrated that pooling is undesirable for tasks that require subtle signals because pooling reinforces content and suppresses noise-like signals \citep{boroumand_deep_2019}. Although this behavior is desirable for computer vision tasks, it is inappropriate for forensic tasks because the noise-like low-level feature is an important clue.

  \subsection{Processing Non-JPEG Images}
  Although JPEG is one of the most widely used formats for storing image data, many other formats are commonly used. Non-JPEG images may not contain DCT coefficients or quantization tables required for the DCT stream. CAT-Net permits these images (assuming they are uncompressed, \eg, PNG) to be its inputs. In this case, CAT-Net computes DCT coefficients by applying DCT to RGB pixel values and assumes the quantization table is filled with ones. This straightforward approach can be implemented with little difference by initially compressing the image with JPEG quality 100 without chroma subsampling. Therefore, CAT-Net can also process non-JPEG images.
  
  A forged image with uncompressed image format does not imply that the original image is uncompressed. It only implies that the final forged image is saved without compression. Thus, compression artifacts during the image acquisition may exist, so analyzing DCT coefficients may be advantageous. However, due to the use of DCT coefficients and quantization tables, the DCT stream is not suitable for analyzing forensic clues in images with other compression formats (\eg, HEIC). In this case, CAT-Net must rely on the RGB stream.

  \section{Double JPEG Detection}
  \label{sec:cat_djpeg}
  
  \begin{figure}[t]
  \centering{\includegraphics[width=1.0\linewidth]{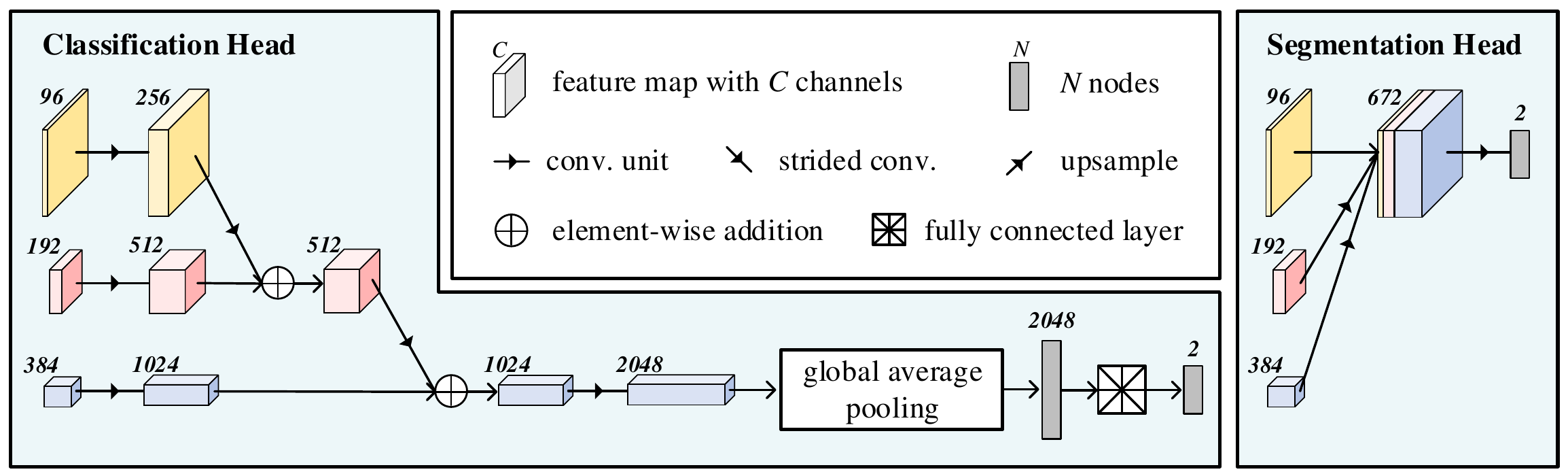}}
  \caption{DCT stream classification and segmentation head architecture. Each is attached at the end of the DCT stream to classify double JPEG images for pretraining (Sect.~\ref{sec:cat_djpeg}) and localize the forgery using the DCT stream for the ablation study (Sect.~\ref{subsec:cat_abl}), respectively. RGB stream heads used for ablation study can be similarly constructed using four resolutions.}
  \label{fig:cat_classification_head}
  \end{figure}
  
  Double JPEG detection is a binary classification task to determine whether a given JPEG image is JPEG compressed once or twice.
  This task requires the ability to analyze the compression artifacts in an image.
  Therefore, the DCT stream, a sub-network of CAT-Net, is pretrained on this task to capture rich compression artifacts.
  The primary purpose of this task is to initialize the image manipulation detection network more efficiently. 
  A classification head is attached at the end to convert the segmentation network to a classification network (Fig.~\ref{fig:cat_classification_head}).

  \subsection{Datasets}
  We used 1.054 million singly and doubly-compressed JPEG images provided by \cite{park_double_2018}.
  They compressed raw images \citep{gloe_dresden_2010, bas_break_2011, dang-nguyen_raise_2015} with 1,120 distinct quantization tables including 51 standard tables (Q50-Q100) and additional custom tables obtained from requested images to their public forensic web service.
  Consequently, their dataset closely represents real-world compression parameters. We used 21 thousand images for testing and the rest for training.
  
  \subsection{Evaluation Metrics}
  Because this is a binary classification task, accuracy (Acc), true positive rate (TPR), and true negative rate (TNR) are measured. We treat doubly compressed images as positives.
  
  \begin{align}
  \textrm{Acc} = \frac{\textrm{TP}+\textrm{TN}}{\textrm{TP}+\textrm{TN}+\textrm{FP}+\textrm{FN}},
  \end{align}
  \begin{align}
  \textrm{TPR} = \frac{\textrm{TP}}{\textrm{TP}+\textrm{FN}}, \: \textrm{TNR} = \frac{\textrm{TN}}{\textrm{TN}+\textrm{FP}}.
  \end{align}
  
  \begin{table}[t]
  \caption{Double JPEG detection performance comparison $(\%)$. Our DCT stream had the highest classification accuracy. Accordingly, the DCT stream learns the compression artifacts accurately. Thus, its weights are used as the initial weights in the image manipulation localization task.}
  \centering
  \resizebox{\linewidth}{!}
  {
      \begin{tabular}{lllll}\\
      \hline
      Method & Input Type & Acc & TPR & TNR\\
      \hline
      ResNet152 & RGB pixels & 54.08 & 0.00 & 100.00\\
      HRNet & RGB pixels & 54.08 & 0.00 & 100.00\\
      ManTra-Net IMTFE & RGB pixels & 54.08 & 0.00 & 100.00\\
      SRNet & RGB pixels & 54.08 & 0.00 & 100.00\\ \hline
      ResNet152 & raw DCT & 54.08 & 0.00 & 100.00\\
      HRNet & raw DCT & 54.08 & 0.00 & 100.00\\ \hline
      \cite{wang_double_2016} & DCT hist. [-5, 5] & 73.05 & 67.74 & 78.37\\ 
      \cite{barni_aligned_2017} & DCT hist. [-60, 60] & 84.46 & 78.35 & 90.53\\ 
      \cite{park_double_2018} & DCT hist. [-60, 60] + QT & 92.76 & 90.90 & 94.59\\ \hline
      ResNet152 & DCT vol. [-20, 20] & 90.19 & 81.97 & 97.17\\
      HRNet & DCT vol. [-20, 20] & 91.56 & 84.60 & 97.47\\
      DCT Stream w/o QT & DCT vol. [-20, 20] & 91.71 & 84.97 & 97.42\\ 
      DCT Stream (Proposed) & DCT vol. [-20, 20] + QT & 93.93 & 89.43 & 97.75\\ \hline
      \end{tabular}
  }
  \label{tab:cat_doubleJPEG}
  \end{table}
  
  \subsection{Results}
  Table~\ref{tab:cat_doubleJPEG} illustrates double JPEG detection results. The first mega row presents four methods taking RGB pixels as inputs. Two general computer vision networks, ResNet~\citep{he2016deep} and HRNet~\citep{wang_deep_2020}, cannot learn compression artifacts at all. 
  Neither can the ManTra-Net~\citep{wu_mantra-net_2019} feature extractor part (Image Manipulation Trace Feature Extractor) learn, as reported by the researchers (Fig. 3 of their paper). SRNet~\citep{boroumand_deep_2019} is a steganalysis network designed to trace minute signals, but it cannot learn either. The RGB domain is not suitable for detecting JPEG double compression.
  The next mega row reveals that two general-purpose networks cannot learn the compression artifacts, supporting our previous claim that CNNs cannot learn the compression artifacts when raw DCT coefficients are supplied directly to them.
  
  The third and last mega rows are the methods using the DCT histogram and DCT volume, respectively. The results reveal that the DCT histogram was indeed a suitable feature representing the distribution of DCT coefficients. The DCT volume is also a highly effective representation of compression artifacts. The DCT stream without the quantization table (\textit{DCT Stream w/o QT}) differs from the normal DCT stream in which the quantization table path and concatenation in Fig.~\ref{fig:cat_JPEG} are removed. \textit{DCT Stream w/o QT} (52.6M) has 30\% fewer parameters than HRNet (75.4M) but a higher accuracy. The proposed DCT stream achieved the highest performance among all methods. The results also demonstrate that adding a quantization table to the network increased the ability to analyze the compression artifacts. The full results confirm that the DCT stream is well designed to capture JPEG compression artifacts.

  \begin{table}[t]
  \caption{Forgery datasets used in the experiments (Sect.~\ref{sec:cat_imd}). These consist of nine publicly available datasets and five custom datasets.}
      \centering
      \small
      \begin{tabular} {lllll}
      \hline
      \multicolumn{2}{l}{Dataset} & Images & JPEGs & QTs \\
      \hline
      \multirow{2}{*}{CASIA v2} & auth. & 7,491 & 7,437 & 50 \\
      & tamp. & 5,105 & 2,057 & 7\\
      \multirow{2}{*}{Fantastic Reality} & auth. & 16,592 & 16,592 & 153\\
      & tamp. & 19,423 & 19,423 & 1 \\
      \multirow{2}{*}{IMD2020} & auth. & 414 & 414 & 58\\
      & tamp. & 2,010 & 1,813 & 73 \\
      NC16 SP & tamp. & 288 & 288 & 3 \\
      \multirow{2}{*}{Carvalho}  & auth. & 100 & 0 & - \\
      & tamp. & 100 & 0 & -\\
      \multirow{2}{*}{Columbia} & auth. & 183 & 0 & -\\
      & tamp. & 180 & 0 & -\\
      \multirow{2}{*}{GRIP} & auth. & 80 & 0 & -\\
      & tamp. & 80 & 0 & -\\
      \multirow{2}{*}{CoMoFoD} & auth. & 200 & 0 & -\\
      & tamp. & 200 & 0 & -\\
      \multirow{2}{*}{COVERAGE} & auth. & 100 & 0 & -\\
      & tamp. & 100 & 0 & -\\
      SP COCO & tamp. & 200,000 & 200,000 & 41\\
      CM COCO & tamp. & 200,000 & 200,000 & 41\\
      CM RAISE & tamp. & 200,000 & 200,000 & 41\\
      CM-JPEG RAISE & tamp. & 200,000 & 200,000 & 41\\
      JPEG RAISE & auth. & 24,462 & 24,462 & 41\\
      \hline
      \end{tabular}
      \label{tab:cat_datasets}
  \end{table}

  \begin{figure*}[t]
  \centering
  \subfigure[SP COCO]{
   \includegraphics[width=0.16\linewidth]{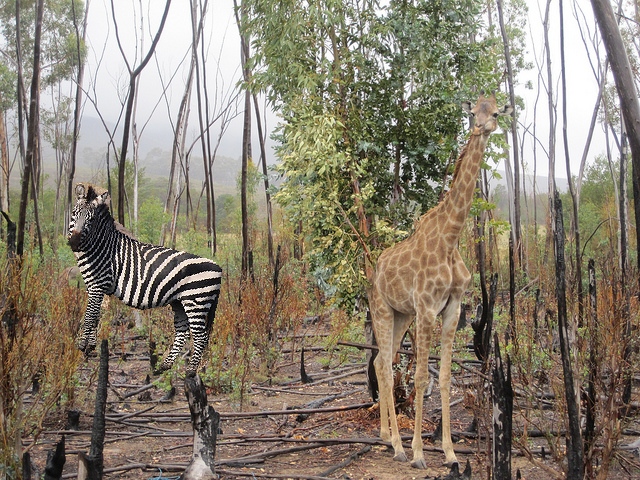}
   \includegraphics[width=0.16\linewidth]{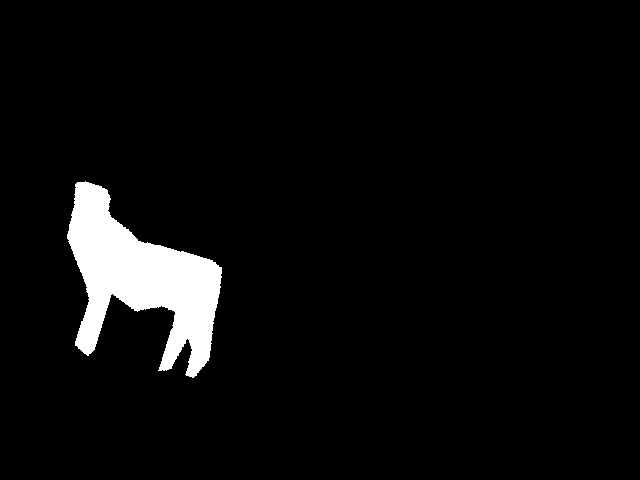}
  }
  \subfigure[CM COCO]{
   \includegraphics[width=0.16\linewidth]{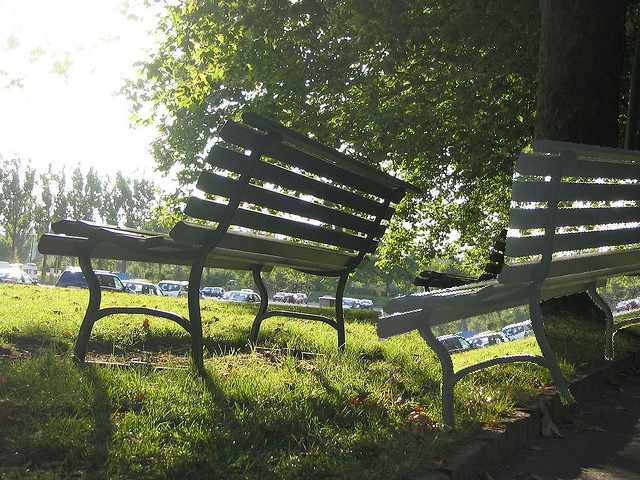}
   \includegraphics[width=0.16\linewidth]{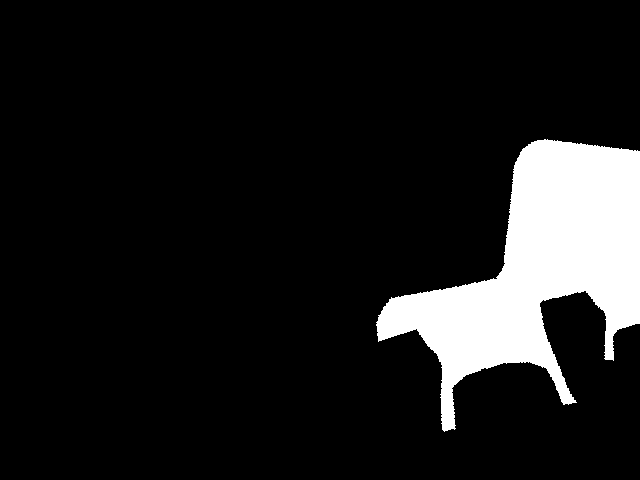}
  }
  \subfigure[CM RAISE]{
   \includegraphics[width=0.12\linewidth]{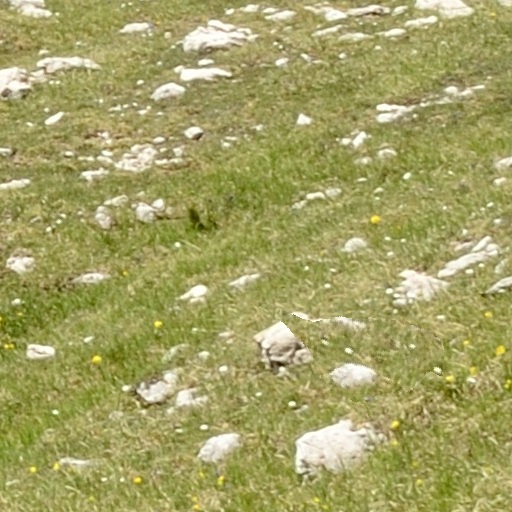}
   \includegraphics[width=0.12\linewidth]{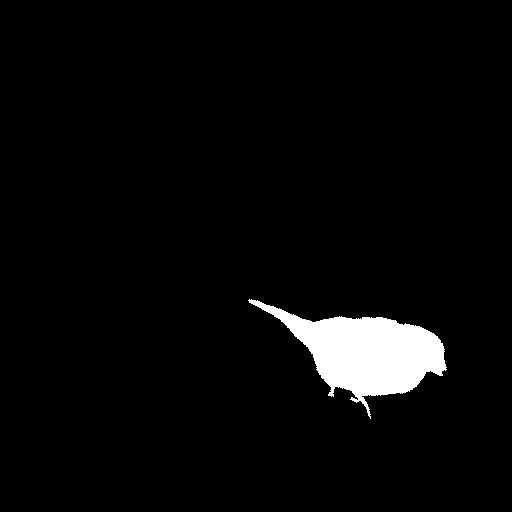}
  }
  \caption{Sample images of created datasets. SP COCO adds an object from another COCO image. CM COCO adds an object from the same COCO image. CM RAISE copies and pastes some random regions. }
  \label{fig:cat_dataset}%
  \end{figure*}

  \subsection{Implementation Details} 
  The third mega row of Table~\ref{tab:cat_doubleJPEG} is from \cite{park_double_2018}, the dataset provider. We performed all other experiments. We used a stochastic gradient descent optimizer with Nesterov momentum (0.9) and weight decay ($10^{-4}$). 
  The learning rate started from 0.05 and decreased by a factor of 0.1 every 10 epochs. We trained until 30 epochs and report the test result for the highest-performing epoch.

  \section{Image Manipulation Detection}
  \label{sec:cat_imd}
  
  This section illustrates the experiments of image manipulation detection and localization.
  After CAT-Net is initialized using ImageNet classification and double JPEG detection, it starts end-to-end training using authentic and tampered images in a supervised manner.
  First, we describe datasets used for training and testing.
  Then, the detailed training methods, baseline methods, and localization evaluation metrics are described.
  We then present the quantitative and qualitative localization results.
  Finally, the ablation studies, effect of compression quality, and robustness tests on additional compression are illustrated.

  \subsection{Datasets} 
  Table~\ref{tab:cat_datasets} summarizes the datasets used in the experiments. We collected nine publicly available datasets.
  \textbf{CASIA v2} \citep{dong_casia_2013} is a popular dataset for copy-move and splicing forgery, images collected from several sources. We used masks provided by a third-party user \citep{pham_hybrid_2019} because ground truth masks are not provided officially.
  \textbf{Fantastic Reality} \citep{kniaz_point_2019} includes many spliced images for various scenes along with ground truth masks. Although authentic images have diverse (153) quantization tables, the tampered images have only one quantization table. 
  \textbf{IMD2020} \citep{novozamsky_imd2020_2020} includes real-life manipulated images and manually created ground truth masks. This dataset contains the most diverse quantization tables because images were collected from the Internet and hence reflect real-world compression schemes.
  \textbf{NC16 SP} \citep{guan_mfc_2019} is a subset of NC16 provided by the National Institutes of Standards and Technology (NIST). NC16 contains high resolution and challenging manipulated images. Although there are several forgery types, we only use splicing forgery.
  \textbf{Carvalho} \citep{carvalho_exposing_2013} DSO-1 contains images of people. Forgeries were created by adding one or more individuals from one image to another with postprocessing to increase photorealism. Blocking artifacts are evident when zoomed in, indicating that although the images are not finally saved in JPEG format, the source images were JPEG compressed, leaving compression artifacts.
  \textbf{Columbia} \citep{ng_data_2004} is a historic dataset for manipulation detection. Ground truth masks are obtained by calculating the difference between authentic and forged images followed by post-processing. The images in this dataset were not compressed in a camera, so they left no compression artifacts.
  \textbf{GRIP} \citep{cozzolino2015efficient} contains realistic copy-move forgery images.
  In that dataset, the ground truth mask contains not only the tampered object region but also the source object region. 
  Thus, we manually remove the source object region so the masks are consistent with the masks in other datasets.
  \textbf{CoMoFoD} \citep{tralic2013comofod} contains copy-move forgeries carefully designed to make forgery detection challenging. The original images were obtained in an uncompressed format. The forged images were heavily postprocessed with JPEG compression, noise adding, image blurring, brightness change, color reduction, and contrast adjustments to hide tampering traces. Manual mask removal of the source object region was also performed on this dataset.
  \textbf{COVERAGE} \citep{wen2016coverage} contains copy-move images designed to counter a similarity-based copy-move forgery detector. Similar but genuine objects are included deliberately, inducing many false positives for those detectors.
  
  We distinguished training and test datasets to measure the ability to generalize over real-world data, \ie, not splitting the same dataset into training and test components. We used the six smallest datasets for testing and the remaining three for training. Those three datasets contain a limited number of images and limited kinds of quantization tables, insufficient to represent real-world image distribution and compression artifacts. Thus, we created five custom datasets and used them for training (Fig.~\ref{fig:cat_dataset}).
  \textbf{SP COCO} was constructed using the COCO 2017 dataset~\citep{lin_microsoft_2015}. Similar to~\cite{wu2018image} and ~\cite{zhou_learning_2018}, spliced images were automatically created by selecting one or more arbitrary objects in one image and pasting them onto another image at random positions, with random rotation and resizing. These images were then compressed. In this paragraph, \textit{compression} refers to JPEG compression at random quality factor ranges from 60–100. We did not apply additional postprocessing, such as blurring the spliced boundary, because that might mislead the network to act like a blur detector.
  \textbf{CM COCO} was constructed similarly, but the copied objects came from the same image. \textbf{CM RAISE} was constructed using RAISE~\citep{dang-nguyen_raise_2015} as an image source but COCO as an object mask. First, the RAISE image was compressed. Then, an arbitrary region was selected using unrelated random polygon annotation from COCO. That region was then pasted within the same image, and finally, the whole image was compressed. This process often creates removal-like forgeries when the background region is selected and copy-pasted. \textbf{CM-JPEG RAISE} was constructed by simply applying additional compression to CM RAISE. This approach mimics the scenario where a forged image is sent through SNS, inducing one more compression. \textbf{JPEG RAISE} is an authentic dataset, created by simply compressing RAISE.

  \subsection{Implementation Details} 
  We initialized CAT-Net weights by pretraining on ImageNet~\citep{krizhevsky_imagenet_2012} classification for the RGB stream and double JPEG classification for the DCT stream (Sect.~\ref{sec:cat_djpeg}). The network was trained end-to-end with authentic and tampered image data.
  We sampled the balanced number of images in each dataset to construct one epoch and efficiently manage the wide variety of dataset sizes. Accordingly, each epoch did not include all training images but only a subset of them. Training images were cropped to $512\times512$ patches aligned with an $8\times8$ grid (Sect.~\ref{sucsec:cat_grid}). Full-resolution images were used for testing, which was possible because the proposed network was fully convolutional.
  The network was implemented in PyTorch~\citep{paszke_pytorch_2019} using a stochastic gradient descent optimizer with a momentum of 0.9. The batch size was 22. We trained for 200 epochs. The learning rate started from 0.005 and decayed exponentially to 0 at the end. The objective was to minimize the pixel-wise binary cross-entropy loss with fivefold more weight on the tampered class. The experiments were performed using two NVIDIA TITAN RTX graphic cards. 
  
  We compared our model performance with ten other methods. 
  The code for seven traditional methods was obtained from MKLab~\citep{zampoglou2017large}: \textbf{DBA}~\citep{ye2007detecting}, \textbf{NOI1}~\citep{mahdian2009using}, \textbf{ADQ}~\citep{lin2009fast}, \textbf{NADQ}~\citep{bianchi2012image}, \textbf{CFA}~\citep{ferrara2012image}, \textbf{NOI2}~\citep{lyu2014exposing}, and \textbf{CAGI}~\citep{iakovidou2018content}. We converted output maps in the range [0, 255] to probability maps in the range [0, 1].
  The code for three deep neural networks and the trained weights were obtained from official public repositories: \textbf{EXIF-SC}~\citep{huh_fighting_2018}, \textbf{ManTra-Net} ~\citep{wu_mantra-net_2019}, and \textbf{Noiseprint}~\citep{cozzolino2019noiseprint}.
  For EXIF-SC, mean-shift was used for output aggregation. ManTra-Net could not infer some extra-large NC16 SP images with full resolution due to GPU memory constraints (24GB). We cropped these 268 images and their corresponding ground truth images to $2560 \times 1440$ (QHD) to test ManTra-Net.

  
  \subsection{Evaluation Metrics}
  Our task is a binary segmentation, labeling each pixel in the input image as tampered (positive, 1) or authentic (negative, 0). Thus, each output pixel can be marked as true positive ($G$:1, $P$:1), true negative ($G$:0, $P$:0), false positive ($G$:0, $P$:1), or false negative ($G$:1, $P$:0), where $G$ is the ground truth mask and $P$ is the prediction output. $G$ and $P$ are 2D binary arrays with the same size as the input image.
  
  We evaluate network performance using accuracy (Acc), F1 score, and average precision (AP).
  The accuracy is defined as:
  \begin{align}
  \textrm{Acc}(G,P) 
  &= \frac{\textrm{TP}+\textrm{TN}}{\textrm{TP}+\textrm{TN}+\textrm{FP}+\textrm{FN}}.
  \end{align}
  However, the problem of accuracy in forensics is that there are much more negative (authentic) pixels than positive (tampered) pixels in the ground truth image. Thus, outputting all pixels as negative produces high accuracy. The F1 score is used to emphasize the positive class --- it is the harmonic mean of precision and recall:
  \begin{align}
      \textrm{F1}(G,P) = \frac{2}{\frac{\textrm{TP}}{\textrm{TP+FP}}+\frac{\textrm{TP}}{\textrm{TP+FN}}} =\frac{2\textrm{TP}}{2\textrm{TP}+\textrm{FP}+\textrm{FN}} .
  \end{align}
  Accuracy and F1 score only measure the binary decision map with a fixed threshold. Although the fixed threshold is indeed essential, we also use average precision to measure the performance free from the threshold. The average precision is the area under the precision-recall curve, which measures an average performance among all thresholds.
  
  For forgery localization tasks, it is sometimes ambiguous which of the two segments is tampered with. Thus, based on \cite{huh_fighting_2018}, we also use the permuted metrics for evaluation, defined as: 
  \begin{align} 
  & \textrm{p-Acc}(G,P) = \textrm{max} \big( \textrm{Acc}(G,P), \textrm{Acc}(G,P ^\complement) \big)
  \\
  & \textrm{p-F1}(G,P) = \textrm{max} \big( \textrm{F1}(G,P), \textrm{F1}(G,P ^\complement) \big)
  \\
  & \textrm{p-AP}(G,P) = \textrm{max} \big( \textrm{AP}(G,P), \textrm{AP}(G,P ^\complement) \big)
  \end{align}
  where $\complement$ negates (flips) the prediction.
  Permuted metrics measure how well a model can distinguish authentic and tampered regions, not its ability to identify which is which.
  Some studies use permuted metrics without explicitly specifying `p-'~\citep{cozzolino2019noiseprint} or use a similar flipping strategy depending on ground truth~\citep{wu_mantra-net_2019} or prediction~\citep{huh_fighting_2018}. Furthermore, some papers use varying thresholds per image and report the best value, resulting in much higher numbers~\citep{cozzolino2019noiseprint, huh_fighting_2018}. The authors claim a varying threshold is used to measure the performance without threshold selection ability. However, because that performance is measured via AP, we chose to use the fixed threshold (0.5) for accuracy and F1 score, to strictly measure the detection performance.
  Each metric is calculated per image and averaged over a dataset.

  \begin{table*}[!p]
      \centering
      \caption{Image manipulation detection and localization performance for completely unseen datasets $(\%)$. Among eleven localization methods, CAT-Net attains the highest localization performance for five out of six forgery datasets in terms of both p-F1 and p-AP. }
      \label{tab:cat_sp}
      \begin{tabular*}{\textwidth}{@{\extracolsep{\fill}}*{9}{l}}
      \hline
      \multicolumn{1}{l}{} & \multicolumn{2}{l}{NC16 SP} & \multicolumn{3}{l}{Carvalho} & \multicolumn{3}{l}{Columbia}  \\
      \cline{2-3} \cline{4-6} \cline{7-9}
      \multicolumn{1}{l}{Method} & \multicolumn{2}{l}{Tamp. (SP)} & \multicolumn{1}{l}{Auth.} & \multicolumn{2}{l}{Tamp. (SP)}& \multicolumn{1}{l}{Auth.} & \multicolumn{2}{l}{Tamp. (SP)} \\
      \cline{2-3} \cline{4-4} \cline{5-6} \cline{7-7} \cline{8-9}
      \multicolumn{1}{l}{}  &  p-F1 & p-AP & p-Acc & p-F1 & p-AP & p-Acc & p-F1 & p-AP  \\
      \hline 
      DBA & 12.60 & 21.13 & 100.00 & 24.48 & 31.87 & 100.00 & 40.87 & 41.48\\ 
      NOI1 & 17.66 & 25.51 & 86.16 & 36.27 & 37.23 & 75.26 & 48.13 & 54.77\\ 
      ADQ & 17.01 & 14.74 & 81.32 & 40.84 & 37.89 & 83.83 & 41.22 & 37.71\\ 
      NADQ & 12.69 & 7.91 & 99.58 & 24.54 & 14.87 & 99.04 & 48.14 & 36.21\\ 
      CFA & 16.59 & 18.60 & 81.98 & 27.87 & 25.88 & 97.52 & 72.54 & 75.02\\ 
      NOI2 & 14.26 & 13.18 & 88.32 & 25.84 & 23.74 & 93.42 & 43.28 & 46.70\\ 
      CAGI & 14.45 & 24.81 & 91.81 & 34.87 & 50.05 & 95.34 & 48.28 & 56.99\\ 
      EXIF-SC & 40.72 & 51.60 & 96.96 & 43.98 & 53.01 & 98.92 & 78.05 & 94.50\\ 
      ManTra-Net & 27.85 & 33.38 & 98.65 & 41.68 & 52.86 & 95.66 & 50.97 & 64.66\\ 
      Noiseprint & 21.51 & 39.89 & 98.58 & 42.12 & 76.79 & 94.07 & 50.42 & 80.85\\ 
      CAT-Net (ours) & 55.62 & 68.76 & 99.91 & 78.79 & 86.41 & 99.61 & 93.97 & 95.87\\ \hline
      \end{tabular*}
  
      \vspace{4mm}
  
      \begin{tabular*}{\textwidth}{@{\extracolsep{\fill}}*{10}{l}}
      \cline{1-10}
      \multicolumn{1}{l}{} & \multicolumn{3}{l}{GRIP} & \multicolumn{3}{l}{CoMoFoD} & \multicolumn{3}{l}{COVERAGE}  \\
      \cline{2-4} \cline{5-7} \cline{8-10}
      \multicolumn{1}{l}{Method} & \multicolumn{1}{l}{Auth.}& \multicolumn{2}{l}{Tamp. (CM)} & \multicolumn{1}{l}{Auth.} & \multicolumn{2}{l}{Tamp. (CM)}& \multicolumn{1}{l}{Auth.} & \multicolumn{2}{l}{Tamp. (CM)} \\
      \cline{2-2} \cline{3-4} \cline{5-5} \cline{6-7} \cline{8-8} \cline{9-10}
      \multicolumn{1}{l}{} & p-Acc & p-F1 & p-AP & p-Acc & p-F1 & p-AP & p-Acc & p-F1 & p-AP  \\
      \hline 
      DBA & 100.00 & 4.24 & 3.78 & 90.24 & 5.32 & 5.76 & 99.94 & 19.57 & 17.80\\ 
      NOI1 & 75.32 & 6.07 & 4.50 & 98.20 & 6.28 & 6.67 & 92.15 & 21.88 & 19.97\\ 
      ADQ & 87.90 & 5.75 & 4.22 & 90.90 & 5.89 & 3.69 & 78.18 & 22.57 & 16.14\\ 
      NADQ & 99.96 & 4.25 & 2.18 & 99.12 & 5.28 & 2.81 & 99.90 & 19.71 & 11.40\\ 
      CFA & 70.64 & 8.81 & 12.83 & 78.85 & 6.33 & 5.71 & 75.16 & 22.93 & 17.23\\ 
      NOI2 & 88.92 & 6.01 & 4.64 & 80.21 & 7.77 & 5.26 & 71.89 & 39.47 & 17.91\\ 
      CAGI & 97.73 & 4.83 & 15.12 & 86.50 & 6.92 & 7.23 & 83.36 & 22.58 & 22.92\\ 
      EXIF-SC & 99.71 & 4.26 & 7.94 & 99.78 & 5.15 & 7.29 & 99.98 & 19.57 & 22.15\\ 
      ManTra-Net & 99.08 & 5.47 & 3.92 & 99.12 & 19.28 & 22.06 & 99.22 & 33.58 & 50.24\\ 
      Noiseprint & 93.32 & 10.98 & 10.23 & 89.87 & 8.99 & 9.33 & 81.09 & 25.46 & 25.45\\ 
      CAT-Net (ours) & 99.46 & 76.45 & 91.87 & 98.68 & 14.01 & 21.46 & 93.00 & 41.27 & 53.76\\ \hline
      \end{tabular*}    
  
      \vspace{5mm}
  
      \caption{Ablation Studies $(\%)$. D.P stands for double JPEG pretraining (Sect.~\ref{sec:cat_djpeg}). }
      \label{tab:cat_ablation_sp}
      \begin{tabular*}{\textwidth}{@{\extracolsep{\fill}}*{9}{l}}
      \cline{1-9}
      \multicolumn{1}{l}{} & \multicolumn{2}{l}{NC16 SP} & \multicolumn{3}{l}{Carvalho} & \multicolumn{3}{l}{Columbia}  \\
      \cline{2-3} \cline{4-6} \cline{7-9}
      \multicolumn{1}{l}{Method} & \multicolumn{2}{l}{Tamp. (SP)} & \multicolumn{1}{l}{Auth.} & \multicolumn{2}{l}{Tamp. (SP)}& \multicolumn{1}{l}{Auth.} & \multicolumn{2}{l}{Tamp. (SP)} \\
      \cline{2-3} \cline{4-4} \cline{5-6} \cline{7-7} \cline{8-9}
      \multicolumn{1}{l}{}  &  p-F1 & p-AP & p-Acc & p-F1 & p-AP & p-Acc & p-F1 & p-AP  \\
      \hline 
      CAT-Net & 55.62 & 68.76 & 99.91 & 78.79 & 86.41 & 99.61 & 93.97 & 95.87\\
      RGB Stream & 42.82 & 55.74 & 99.80 & 40.68 & 61.10 & 99.96 & 94.26 & 96.59\\ 
      DCT Stream & 41.75 & 49.87 & 99.53 & 64.85 & 74.99 & 99.93 & 71.88 & 85.93\\ 
      CAT-Net w/o D.P & 47.43 & 60.76 & 99.90 & 51.33 & 68.87 & 99.98 & 95.15 & 98.21\\ \hline
      \end{tabular*}
  
      \vspace{4mm}
  
      \begin{tabular*}{\textwidth}{@{\extracolsep{\fill}}*{10}{l}}
      \cline{1-10}
      \multicolumn{1}{l}{} & \multicolumn{3}{l}{GRIP} & \multicolumn{3}{l}{CoMoFoD} & \multicolumn{3}{l}{COVERAGE}  \\
      \cline{2-4} \cline{5-7} \cline{8-10}
      \multicolumn{1}{l}{Method} & \multicolumn{1}{l}{Auth.}& \multicolumn{2}{l}{Tamp. (CM)} & \multicolumn{1}{l}{Auth.} & \multicolumn{2}{l}{Tamp. (CM)}& \multicolumn{1}{l}{Auth.} & \multicolumn{2}{l}{Tamp. (CM)} \\
      \cline{2-2} \cline{3-4} \cline{5-5} \cline{6-7} \cline{8-8} \cline{9-10}
      \multicolumn{1}{l}{} & p-Acc & p-F1 & p-AP & p-Acc & p-F1 & p-AP & p-Acc & p-F1 & p-AP  \\
      \hline 
      CAT-Net & 99.46 & 76.45 & 91.87 & 98.68 & 14.01 & 21.46 & 93.00 & 41.27 & 53.76\\ 
      RGB Stream & 99.97 & 9.46 & 17.63 & 99.66 & 17.48 & 32.99 & 92.50 & 43.03 & 54.84\\ 
      DCT Stream & 99.90 & 65.94 & 81.96 & 99.53 & 8.89 & 13.12 & 94.06 & 34.96 & 42.36\\ 
      CAT-Net w/o D.P & 99.97 & 24.60 & 43.43 & 99.66 & 15.53 & 29.95 & 97.84 & 39.46 & 57.04\\ \hline
      \end{tabular*}
  \end{table*}

  \begin{figure*}[!p]
  \centering{\includegraphics[width=1.0\linewidth]{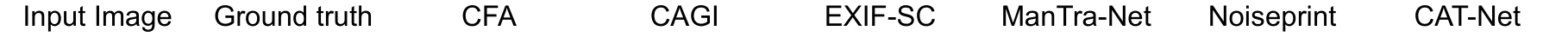}}
  \centering{\includegraphics[width=1.0\linewidth]{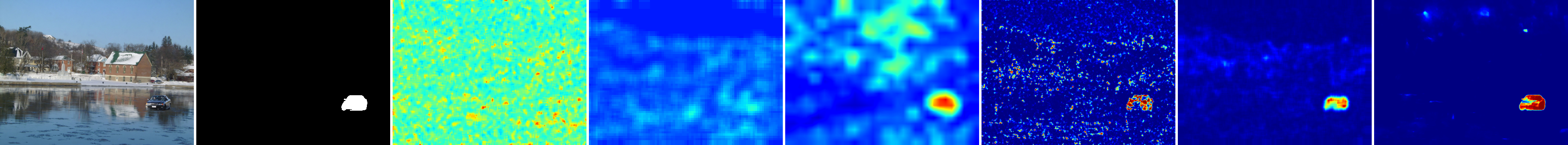}}
  \centering{\includegraphics[width=1.0\linewidth]{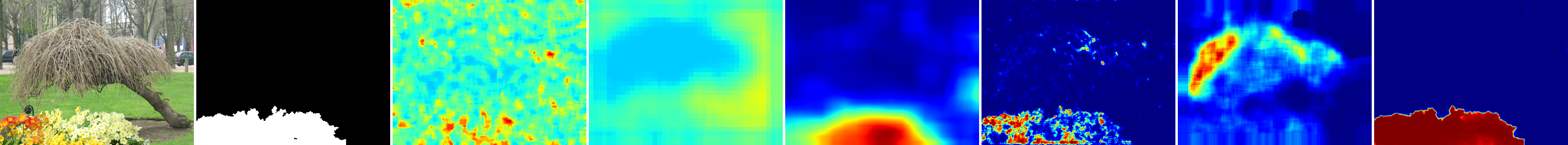}}
  \centering{\includegraphics[width=1.0\linewidth]{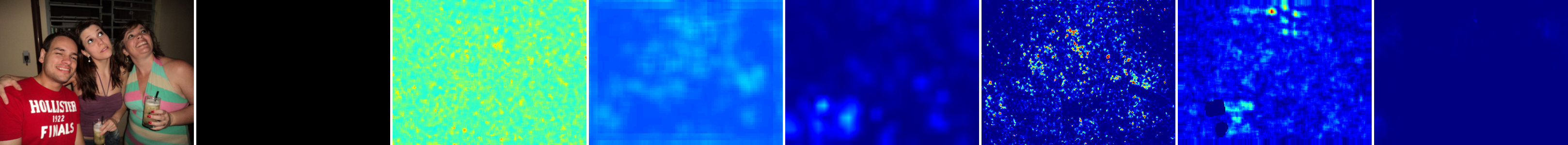}}
  \centering{\includegraphics[width=1.0\linewidth]{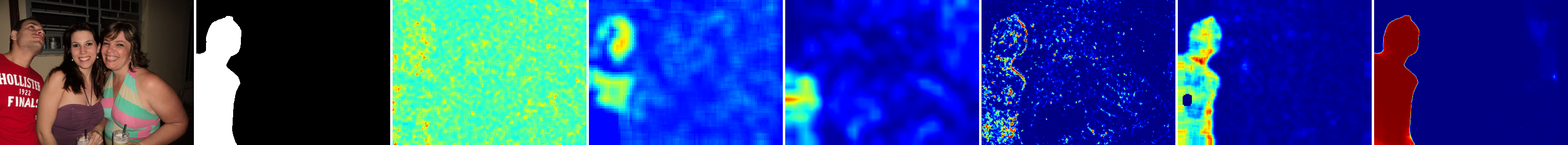}}
  \centering{\includegraphics[width=1.0\linewidth]{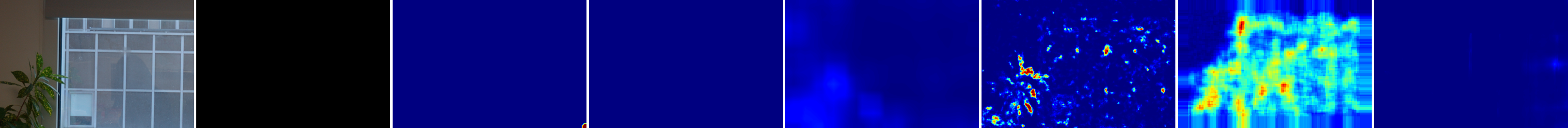}}
  \centering{\includegraphics[width=1.0\linewidth]{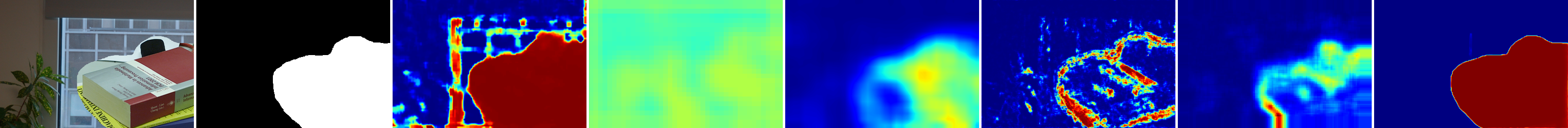}}
  \centering{\includegraphics[width=1.0\linewidth]{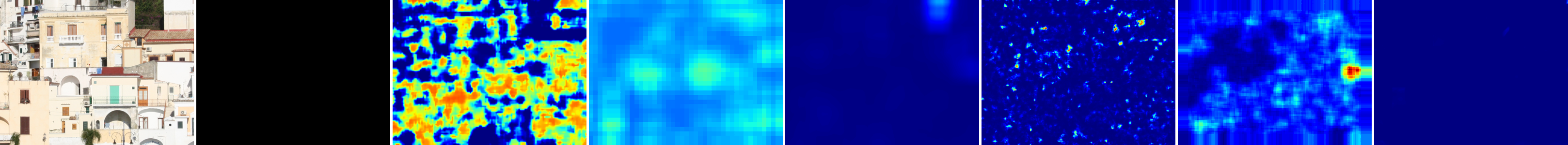}}
  \centering{\includegraphics[width=1.0\linewidth]{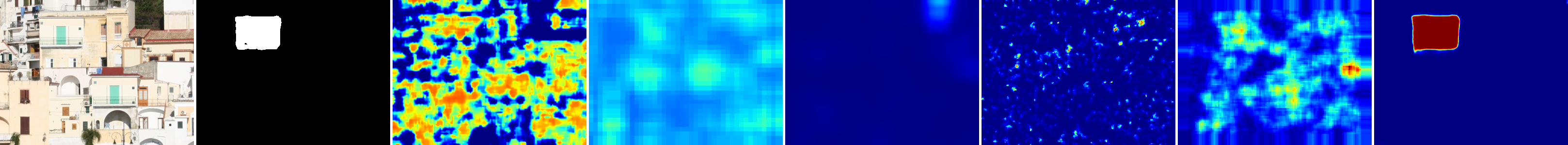}}
  \centering{\includegraphics[width=1.0\linewidth]{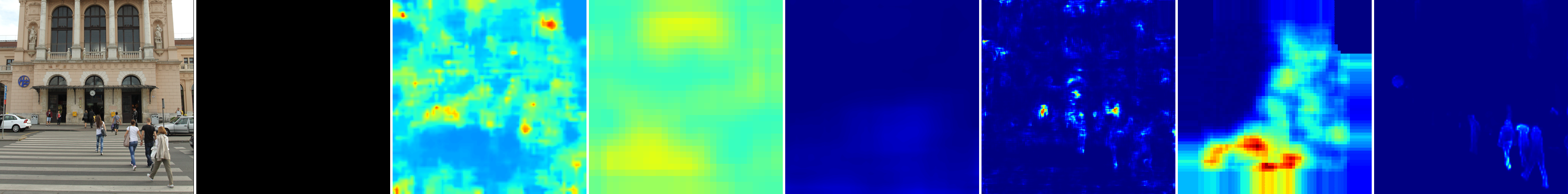}}
  \centering{\includegraphics[width=1.0\linewidth]{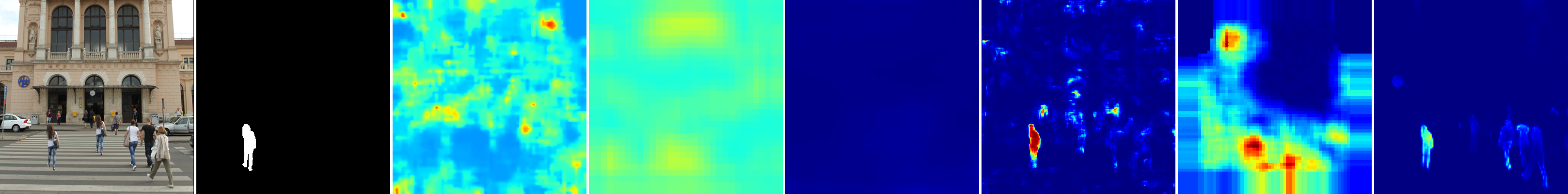}}
  \centering{\includegraphics[width=1.0\linewidth]{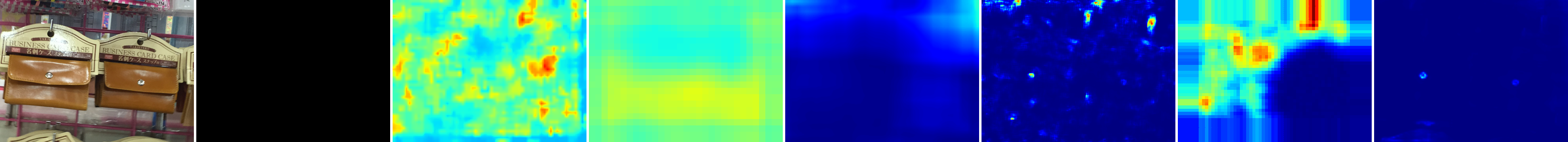}}
  \centering{\includegraphics[width=1.0\linewidth]{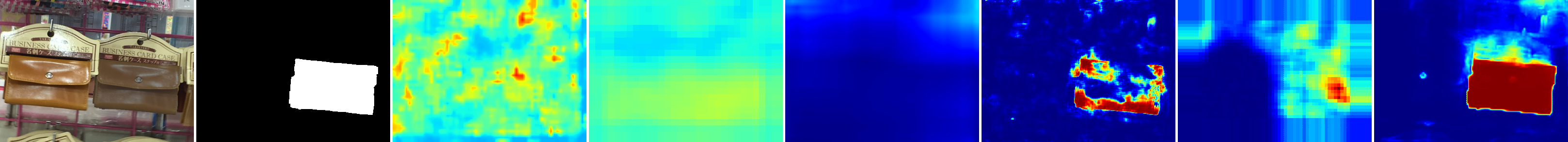}}
  \caption{Image manipulation detection and localization results. The colors indicate  confidence of being tampered with. The color bar is depicted in Fig.~\ref{fig:cat_problem_formulation}. From top to bottom: 2x NC16 SP, Carvalho (auth.), Carvalho (tamp.), Columbia (auth.), Columbia (tamp.), GRIP (auth.), GRIP (tamp.), CoMoFoD (auth.), CoMoFoD (tamp.), COVERAGE (auth.), COVERAGE (tamp.).}
  \label{fig:cat_heatmap_main}
  \end{figure*}


  \subsection{Results}
  Table~\ref{tab:cat_sp} presents a performance comparison among eleven methods: seven traditional approaches, three state-of-the-art deep neural networks, and our CAT-Net. The results are depicted for six independent datasets: three splicing datasets and three copy-move datasets. All test datasets are completely unseen during training,~\ie, not test splits, to measure the general performance for real-world forgeries. We also report accuracy for authentic images in each dataset, if provided, to observe the false positive rate for untampered images. Figure~\ref{fig:cat_heatmap_main} illustrates some prediction outputs of the six highest-performing methods.
  
  Among eleven methods, CAT-Net achieves the highest localization performance for five out of six forgery datasets in terms of both p-F1 and p-AP. In particular, the CAT-Net results for GRIP are surprising (Table~\ref{tab:cat_sp}). CAT-Net achieves 76.45\% p-F1 and 91.87\% p-AP, while the second-best methods are 10.98\% p-F1 (Noiseprint) and 15.12\% p-AP (CAGI). The other ten methods could not localize the forgeries, unlike CAT-Net, which significantly outperformed those methods. Although GRIP creators tried not to leave any forgery traces, the acquisition-level compression artifacts remained and those are detected by our detector. However, CAT-Net could not detect forgeries well in CoMoFoD (14.01\% p-F1, 21.46\% p-AP), defeated by ManTra-Net (19.28\% p-F1, 22.06\% p-AP). The result was caused by the absence of initial compression, which indicates the DCT stream could find no traces. In contrast, although Columbia also does not contain initial compression traces, CAT-Net achieves excellent performance due to the RGB stream. CAT-Net attains 93.97\% p-F1 and 95.87\% p-AP, while the second-highest performing method, EXIF-SC, achieves 78.05\% p-F1 and 94.50\% p-AP. Their method is suitable for Columbia because this dataset is claimed to be uncompressed, so EXIF metadata is unharmed.
  
  The permuted accuracy of authentic images suggests that DBA has the lowest false positives for untampered images. However, because its localization performance for tampered images is low in many cases, we conclude that this method predicts tampered regions relatively less often. CAT-Net achieves a high score for authentic and tampered images, implying that it could be used for image integrity verification. CAT-Net achieves state-of-the-art performance in image manipulation detection and localization.
  
  \begin{figure*}[t]%
  \centering{\includegraphics[width=0.8\linewidth]{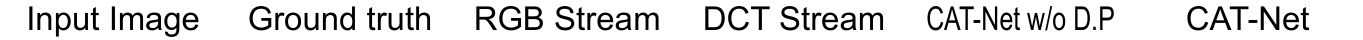}}
  \centering{\includegraphics[width=0.8\linewidth]{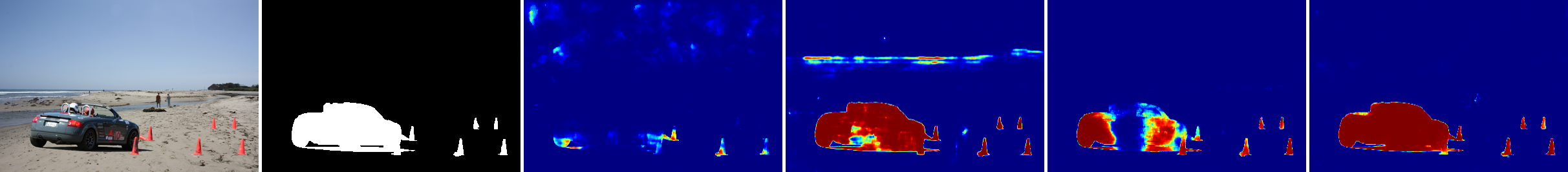}}
  \centering{\includegraphics[width=0.8\linewidth]{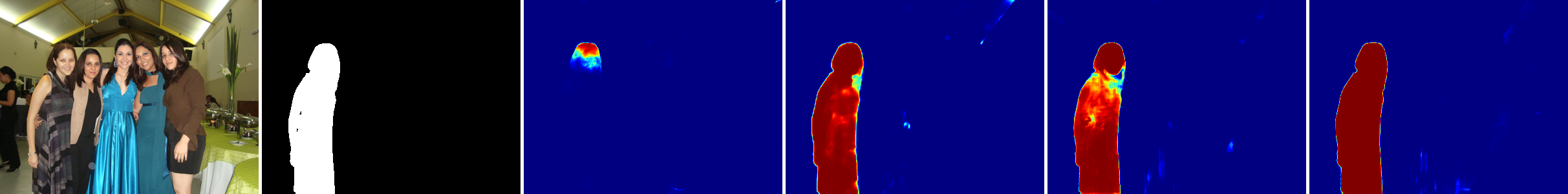}}
  \centering{\includegraphics[width=0.8\linewidth]{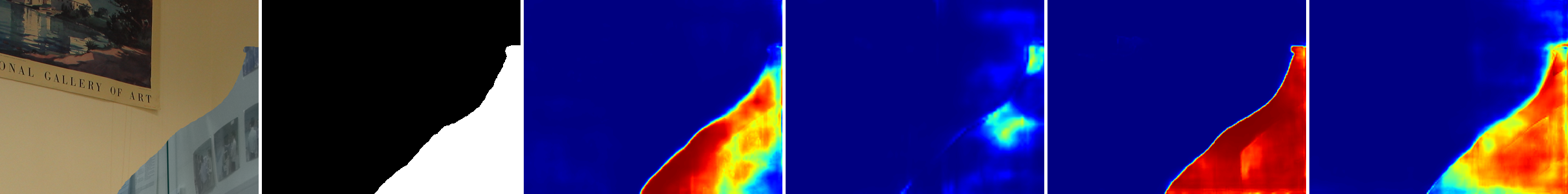}}
  \centering{\includegraphics[width=0.8\linewidth]{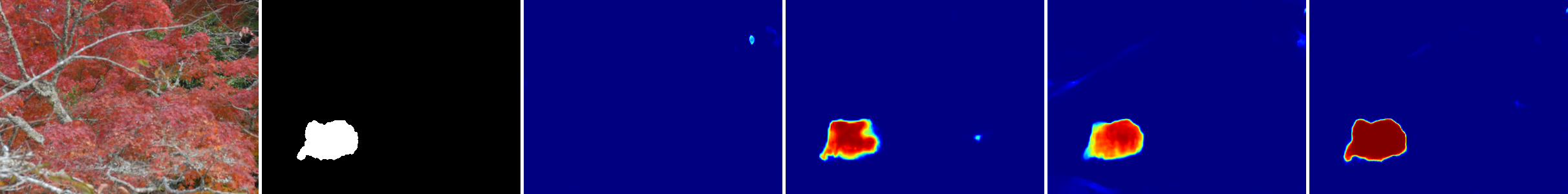}}
  \centering{\includegraphics[width=0.8\linewidth]{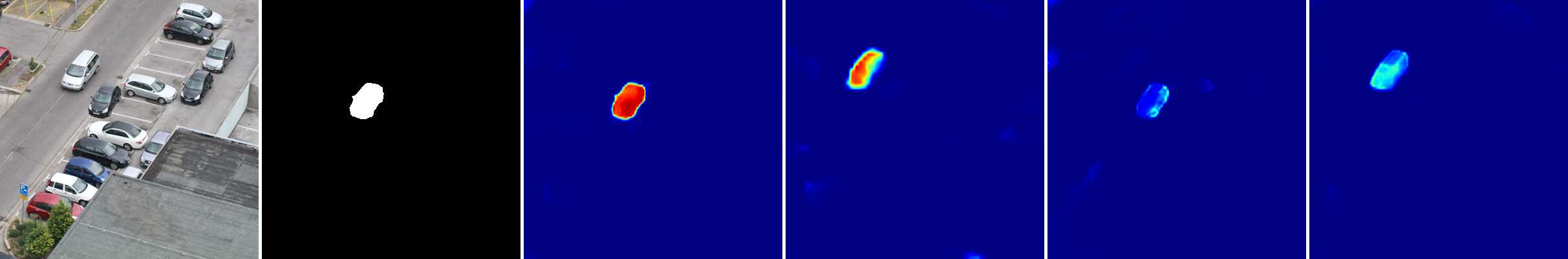}}
  \centering{\includegraphics[width=0.8\linewidth]{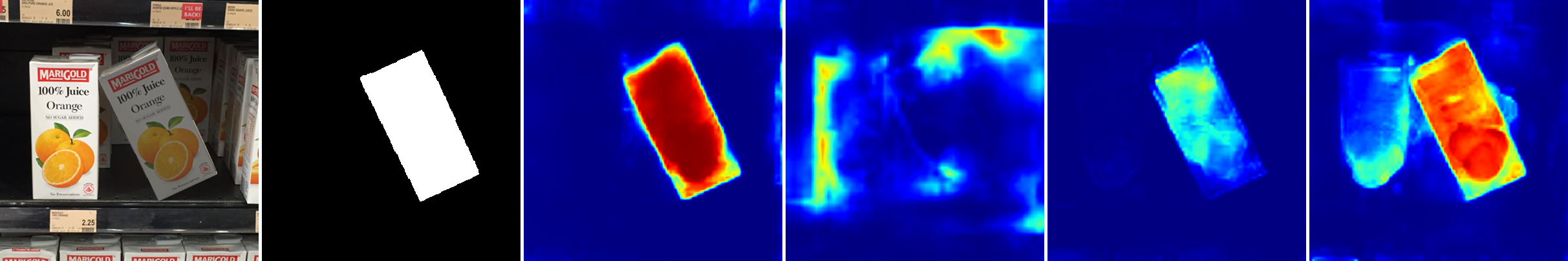}}
  \centering{\includegraphics[width=0.32\linewidth]{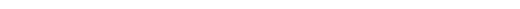}}
  \caption{Ablation studies (Sect.~\ref{subsec:cat_abl}). From top to bottom: NC16 SP, Carvalho (tamp.), Columbia (tamp.), GRIP (tamp.), CoMoFoD (tamp.), COVERAGE (tamp.).}
  \label{fig:cat_heatmap_abl}
  \end{figure*}

  \begin{figure}[t]%
  \centering{
    \includegraphics[width=1.0\linewidth]{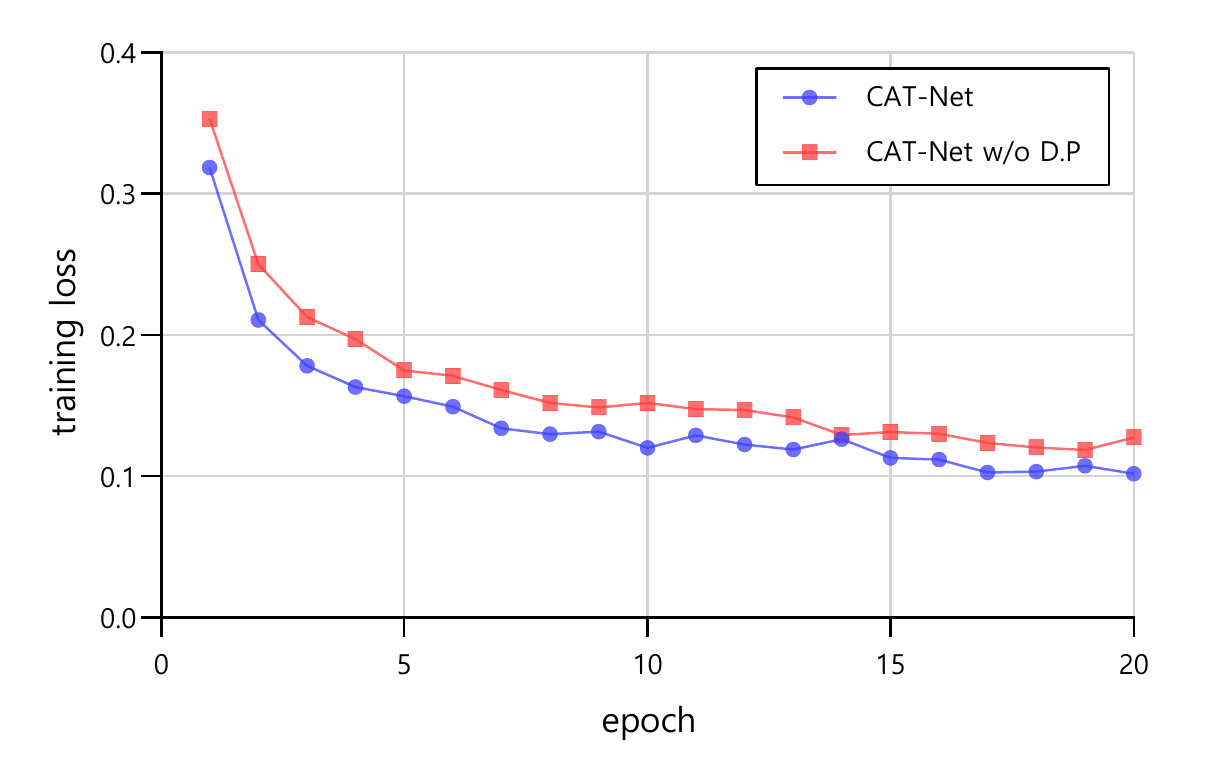}}
  \caption{Effect of double JPEG detection pretraining on the DCT stream. Pretraining on double JPEG detection produced a faster drop in training loss. The first 20 epochs out of 200 epochs are represented.}
  \label{fig:cat_pretrain_graph}
  \end{figure}

  \subsection{Ablation Studies} 
  \label{subsec:cat_abl}

  Table~\ref{tab:cat_ablation_sp} and Fig.~\ref{fig:cat_heatmap_abl} present the ablation study results. Two substreams are separately trained using the same training settings to observe the contribution of each stream. In some cases, the DCT stream outperforms the RGB stream and, in others, the opposite --- depending on the existence of compression artifacts. If the compression artifacts exist, the DCT stream outperforms the RGB stream and vice versa if the DCT stream cannot trace meaningful compression traces. For the datasets without initial compression, the joint performance sometimes decreases because the DCT stream produces unhelpful features, negatively impacting the entire network. Nevertheless, full CAT-Net exhibits the highest overall performance using both streams.
  
  The last row of Table~\ref{tab:cat_ablation_sp} illustrates the effect of double JPEG pretraining. \textit{CAT-Net w/o D.P} indicates CAT-Net started training from random initialization for the DCT stream, not from the pretrained weights using double JPEG detection. 
  For the datasets with compression traces, pretraining on double JPEG detection improves the localization performance significantly. For example, \textit{CAT-Net w/o D.P} attains 24.60\% p-F1 and 43.43\% p-AP for GRIP. The performance increases to 76.45\% p-F1 and 91.87\% p-AP when the training starts from double JPEG initialization. Furthermore, Fig.~\ref{fig:cat_pretrain_graph} illustrates that double JPEG pretraining produces faster training. These results are likely due to the various quantization tables used in double JPEG pretraining (1,120 types). It is challenging to acquire a forged image and a ground truth mask pairs for diverse quantization tables. In contrast, it is easy to obtain singly and doubly-compressed images with diverse quantization tables because we may obtain raw images and compress them once or twice. Therefore, we recommend using double JPEG pretraining for future research. 
  
  However, when tested on the datasets without useful compression traces, the overall performance decreases when pretrained. For example, \textit{CAT-Net w/o D.P} exhibits 95.15\% p-F1 and 98.21\% p-AP for Columbia, which decrease to 93.97\% p-F1 and 95.87\% p-AP when started from double JPEG pretraining. When we use pretraining, the DCT stream produces more accurate predictions during training, causing the entire network to focus more on the DCT stream than the RGB stream. When evaluating images without compression traces, this stream tries to use unavailable compression traces more frequently, reducing performance. 
  Hence, we conclude that pretraining the DCT stream with double JPEG detection enables rich initialization of the forgery localization task where compression artifacts remain.

  \begin{figure*}[t]
  \centering{\includegraphics[width=0.85\linewidth]{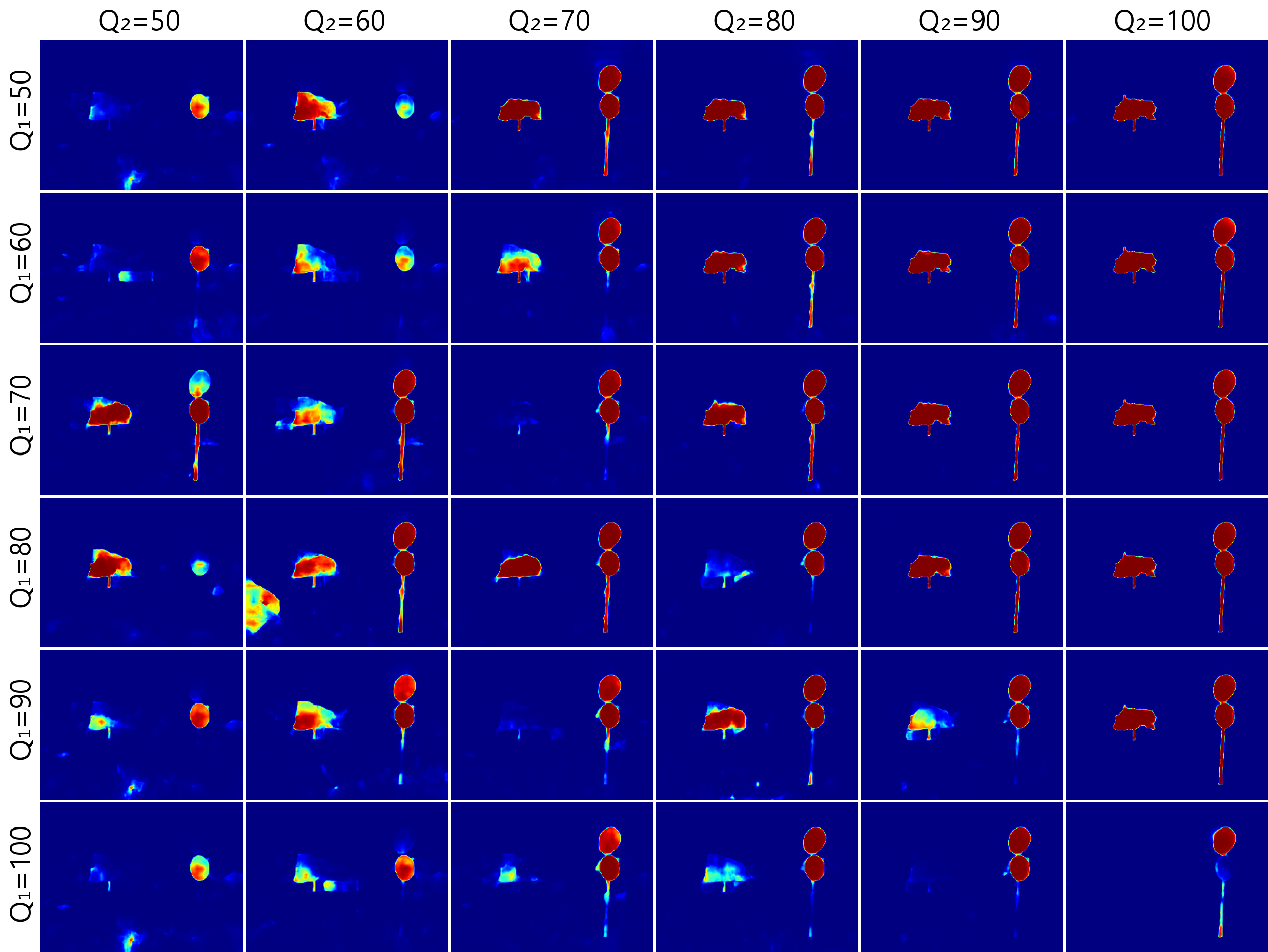}}
  \caption{Effect of first and second compression qualities on localization performance (Sect.~\ref{subsec:cat_quant_analysis}). 
  Forged image and ground truth are depicted in Figs.~\ref{fig:cat_problem_b} and~\ref{fig:cat_problem_c}, respectively.}
  \label{fig:cat_double_comp}
  \end{figure*}

  \begin{figure*}[!p]
  \centering
  \subfigure[NC16 SP, p-F1]{
   \includegraphics[width=0.30\linewidth]{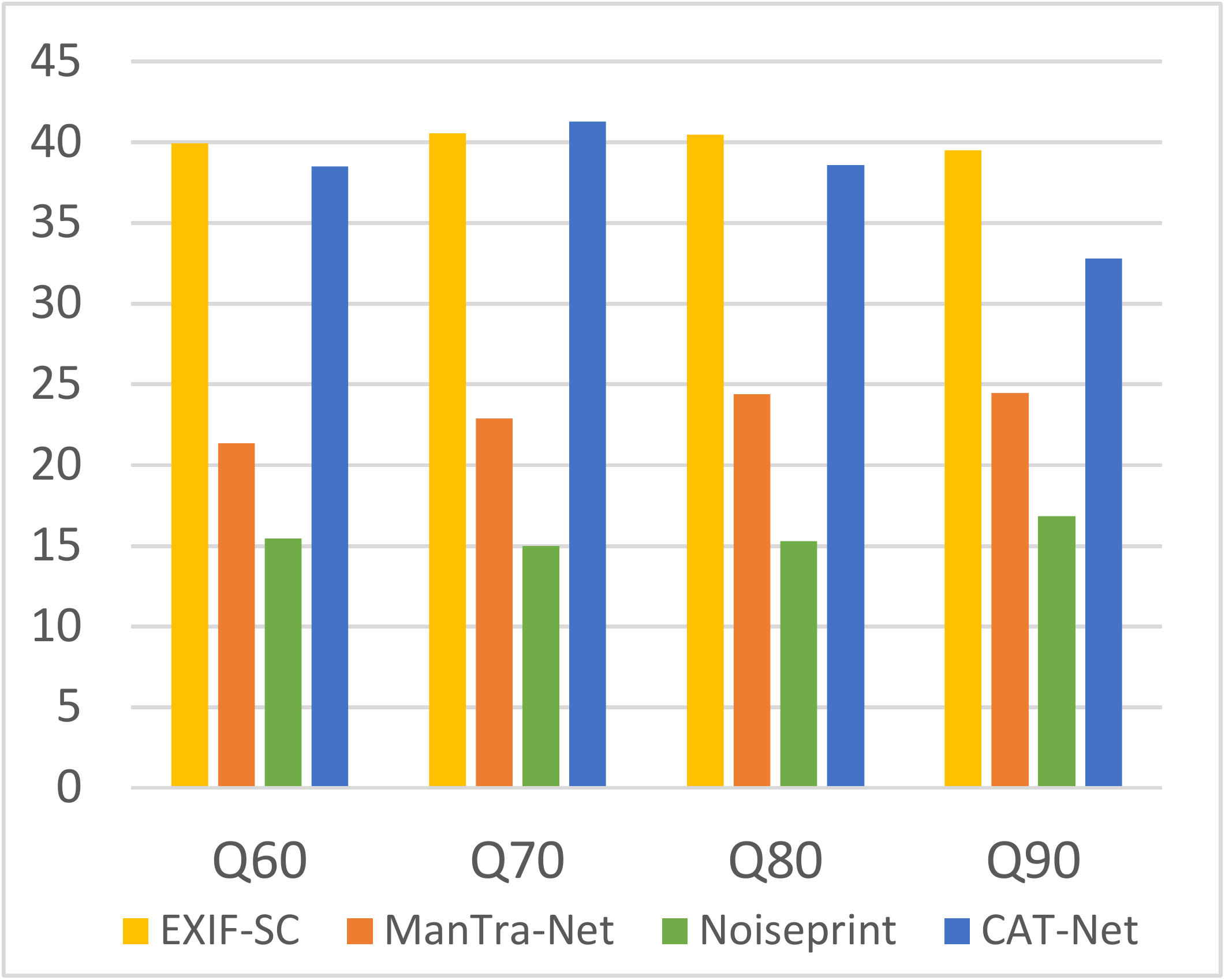}
  }
  \subfigure[Carvalho, p-F1]{
   \includegraphics[width=0.30\linewidth]{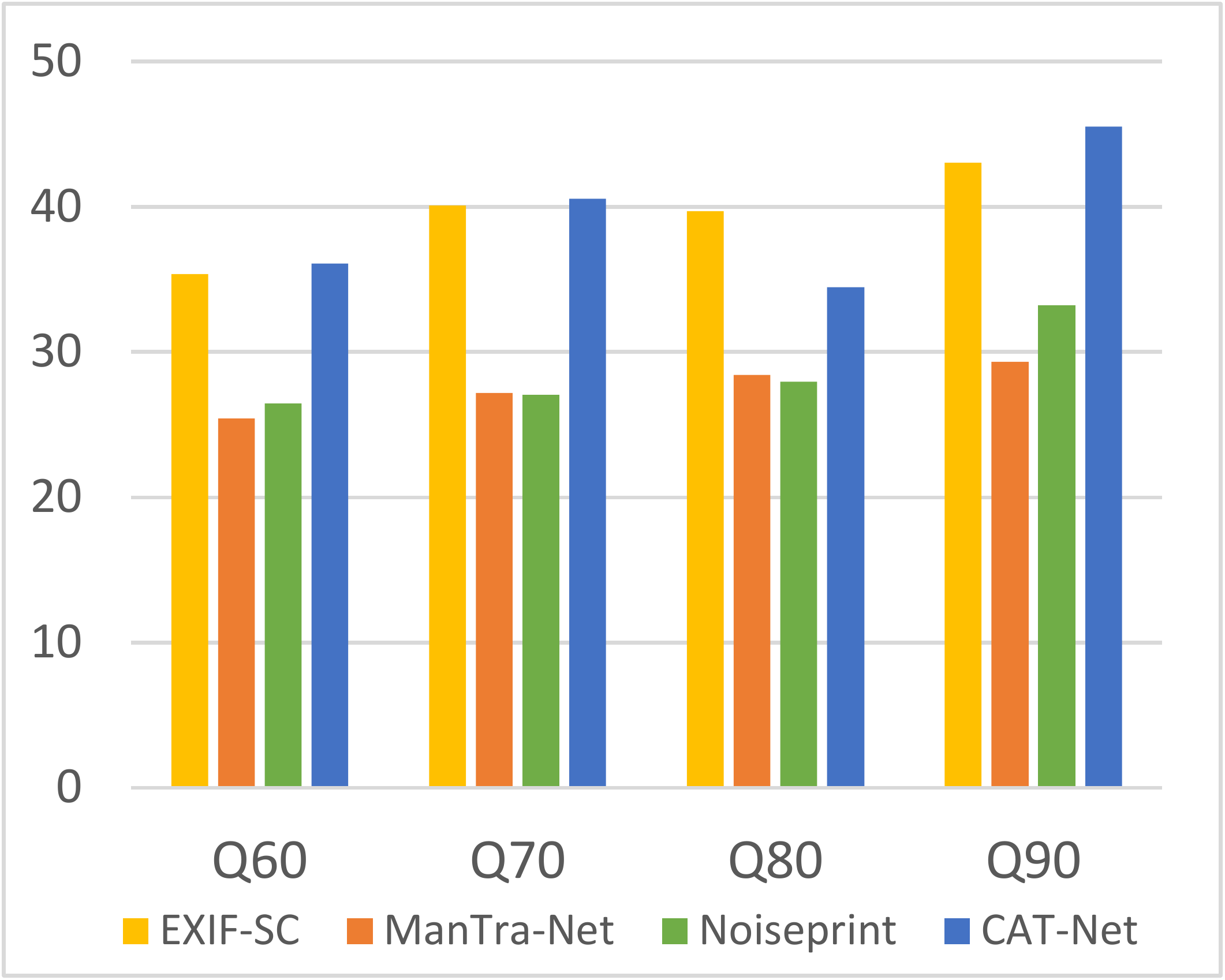}
  }
  \subfigure[Columbia, p-F1]{
   \includegraphics[width=0.30\linewidth]{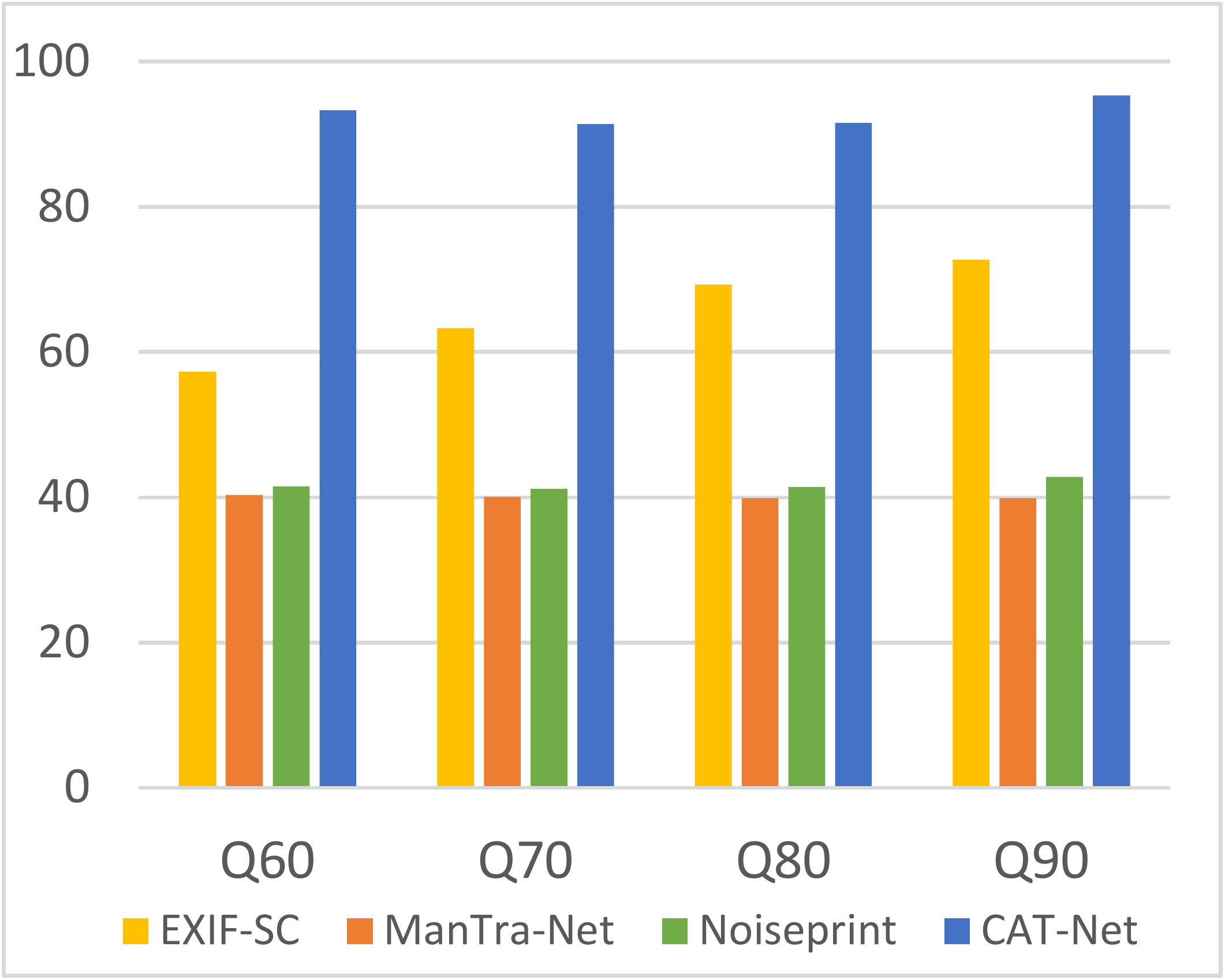}
  }
  \vspace{3mm}
  
  \subfigure[GRIP, p-F1]{
   \includegraphics[width=0.30\linewidth]{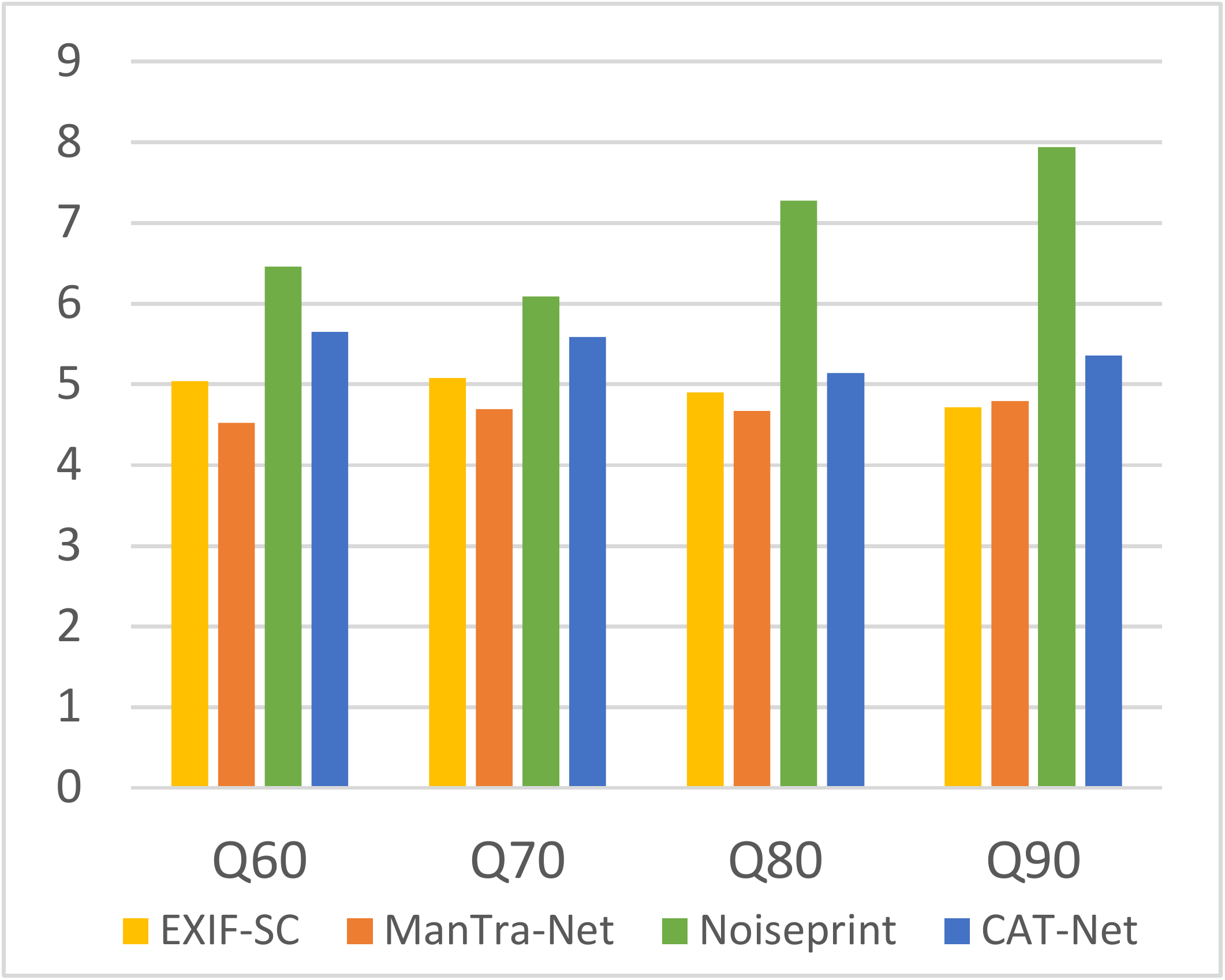}
  }
  \subfigure[CoMoFoD, p-F1]{
   \includegraphics[width=0.30\linewidth]{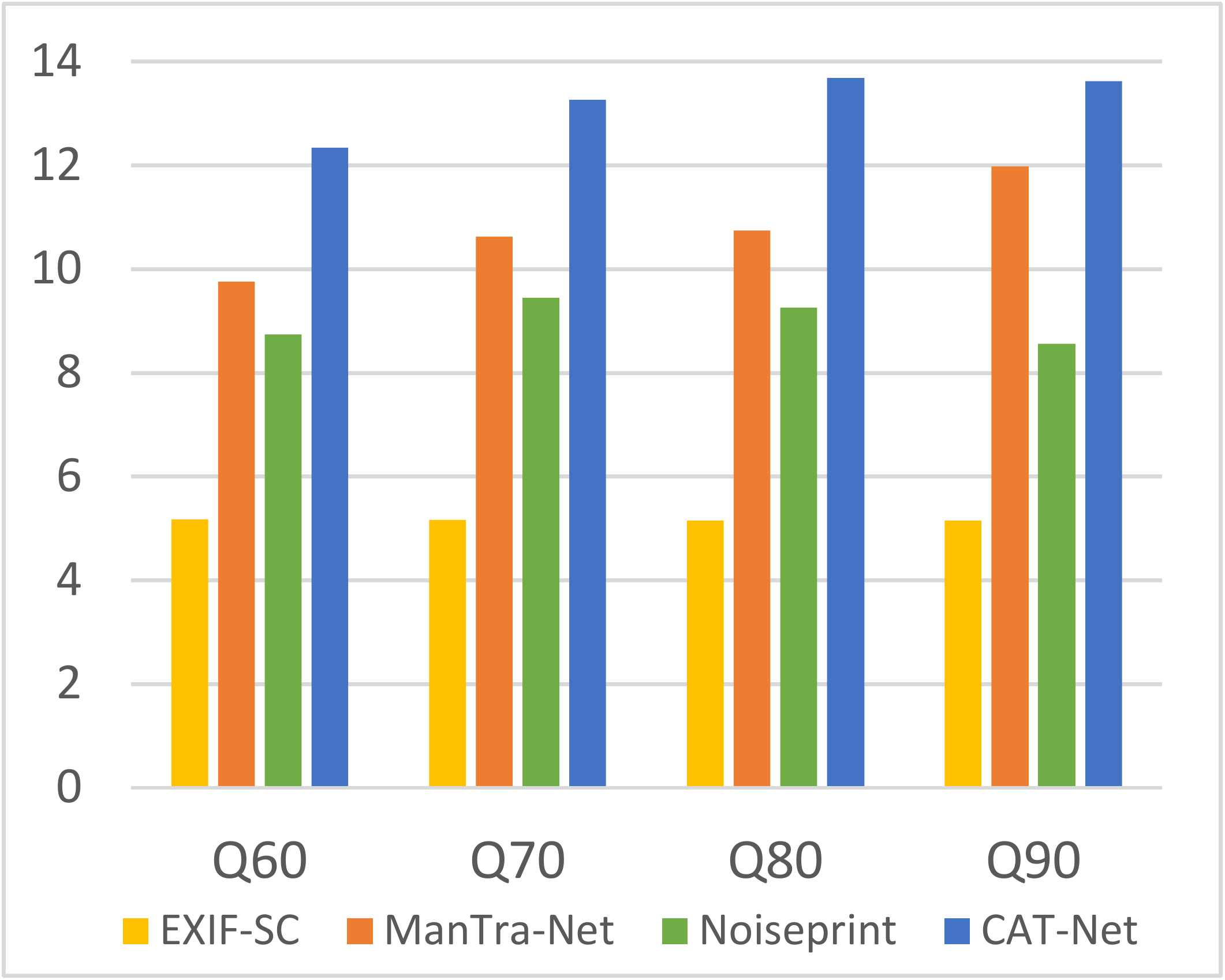}
  }
  \subfigure[COVERAGE, p-F1]{
   \includegraphics[width=0.30\linewidth]{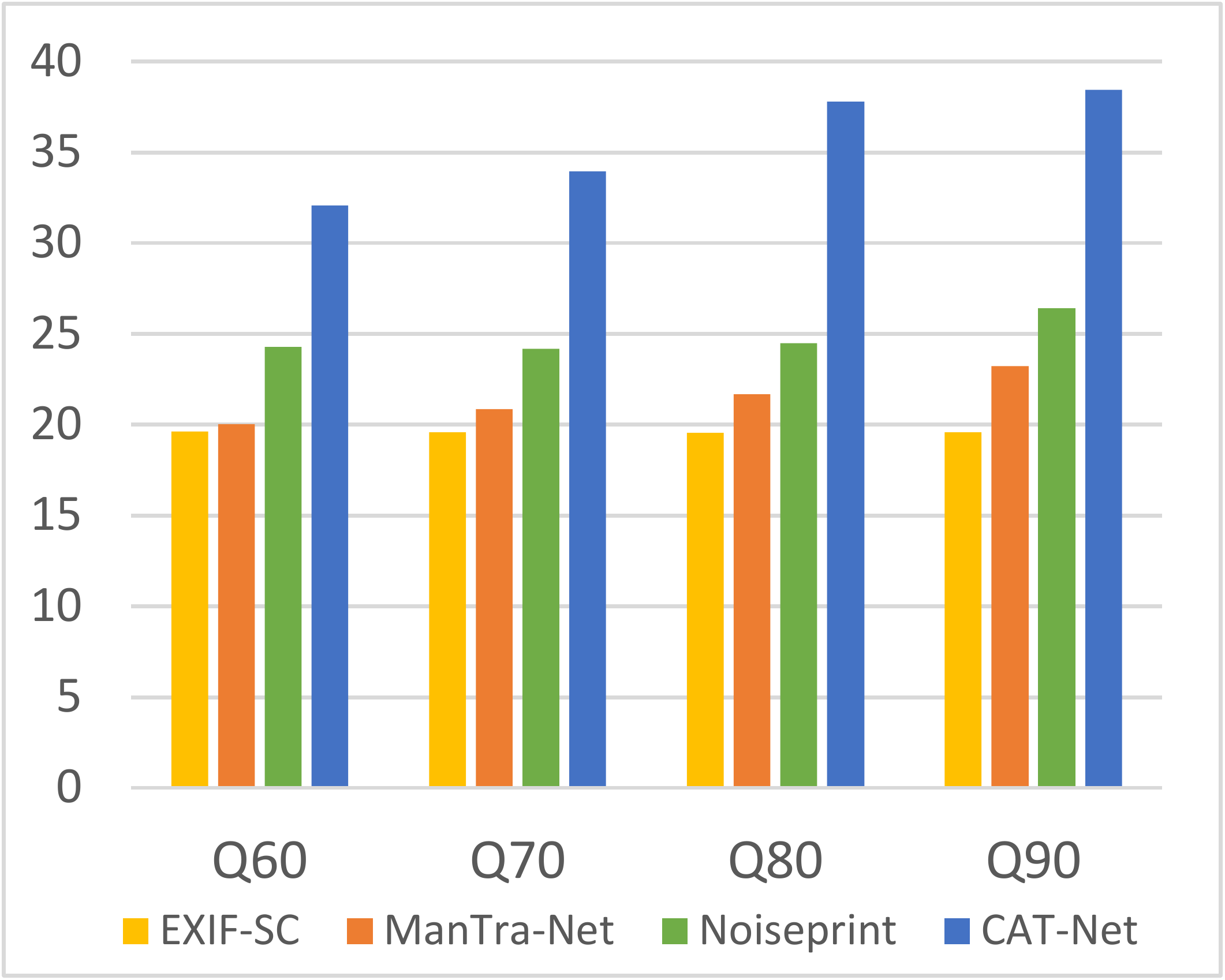}
  }
  \vspace{3mm}
  
  \subfigure[NC16 SP, p-AP]{
   \includegraphics[width=0.30\linewidth]{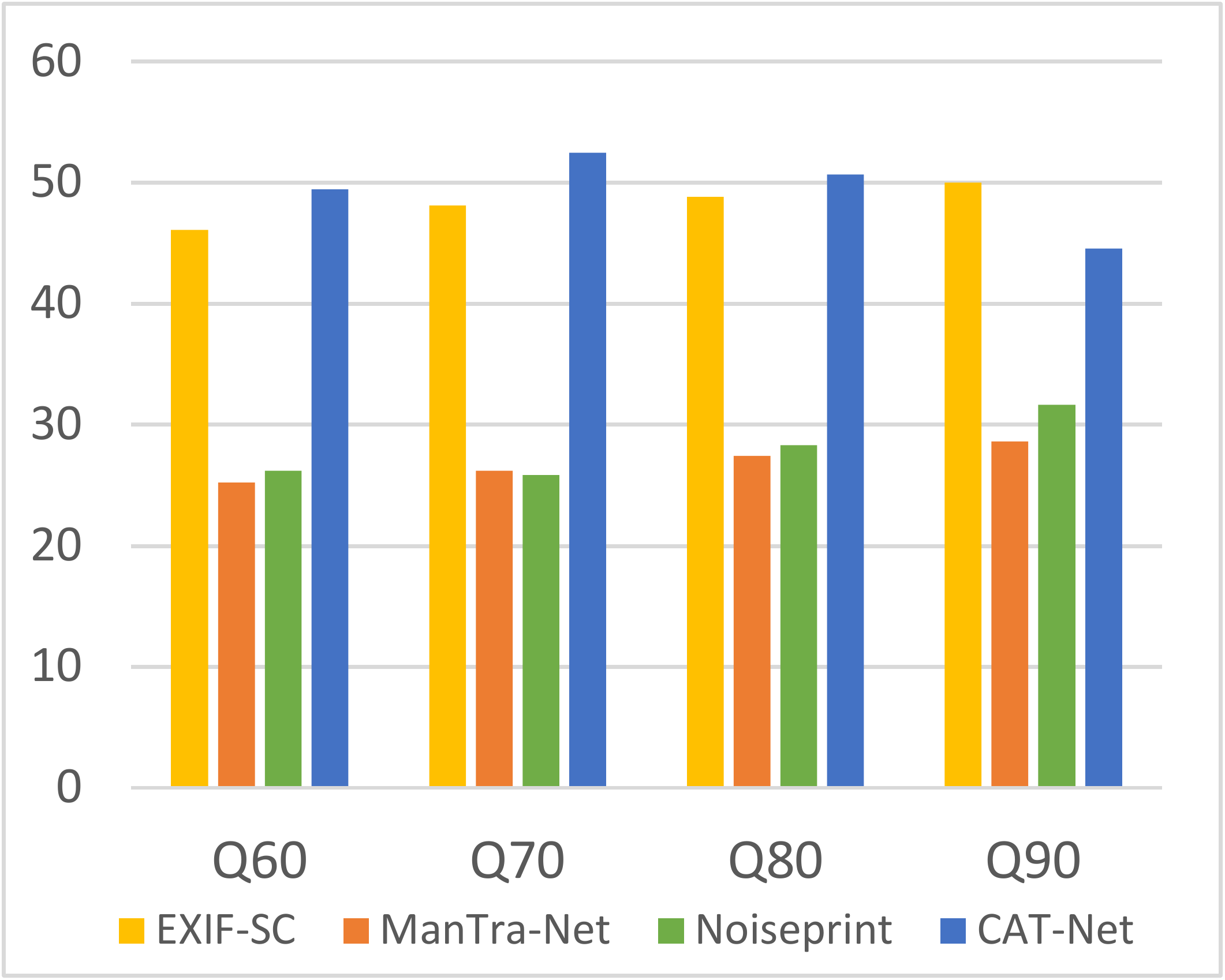}
  }
  \subfigure[Carvalho, p-AP]{
   \includegraphics[width=0.30\linewidth]{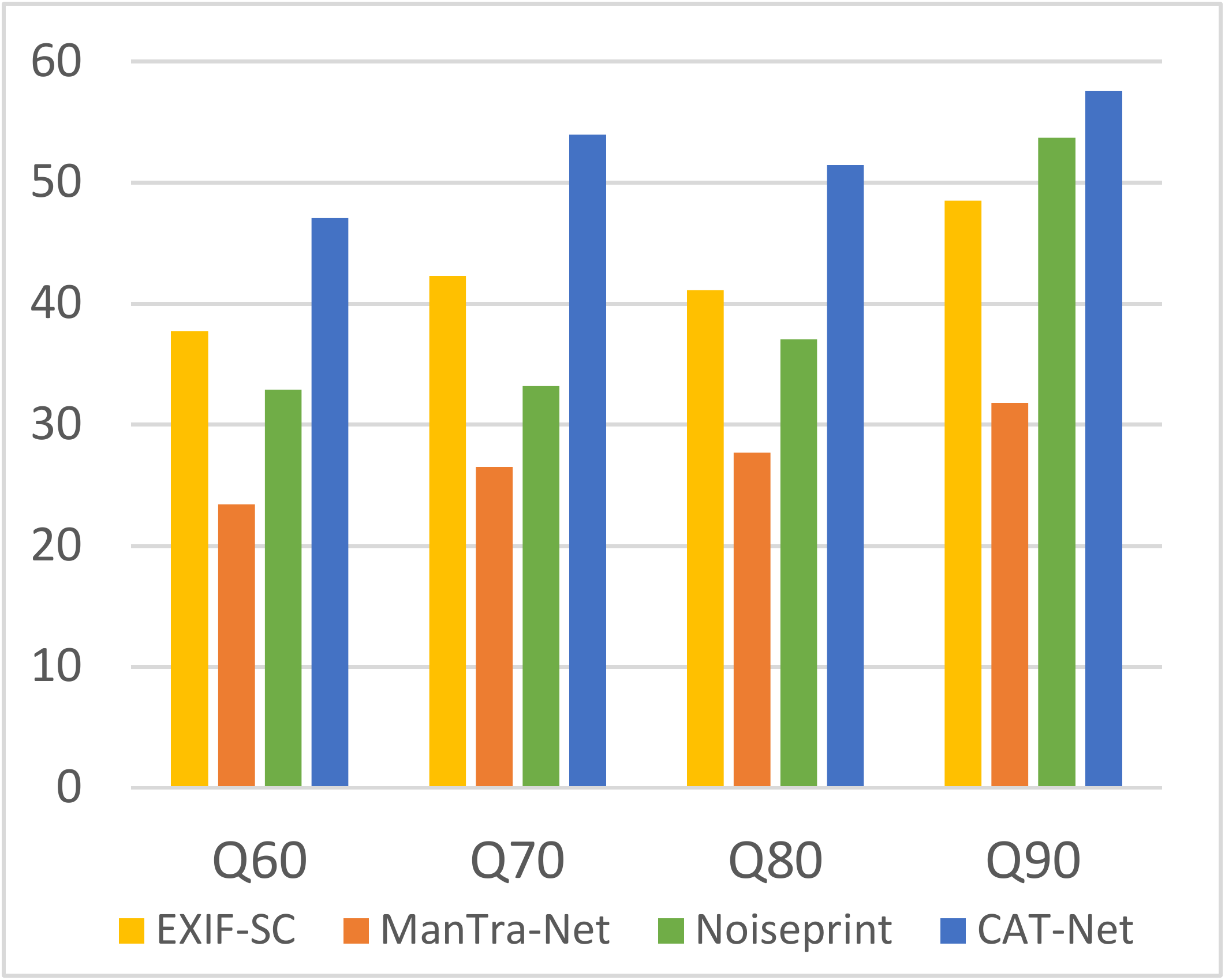}
  }
  \subfigure[Columbia, p-AP]{
   \includegraphics[width=0.30\linewidth]{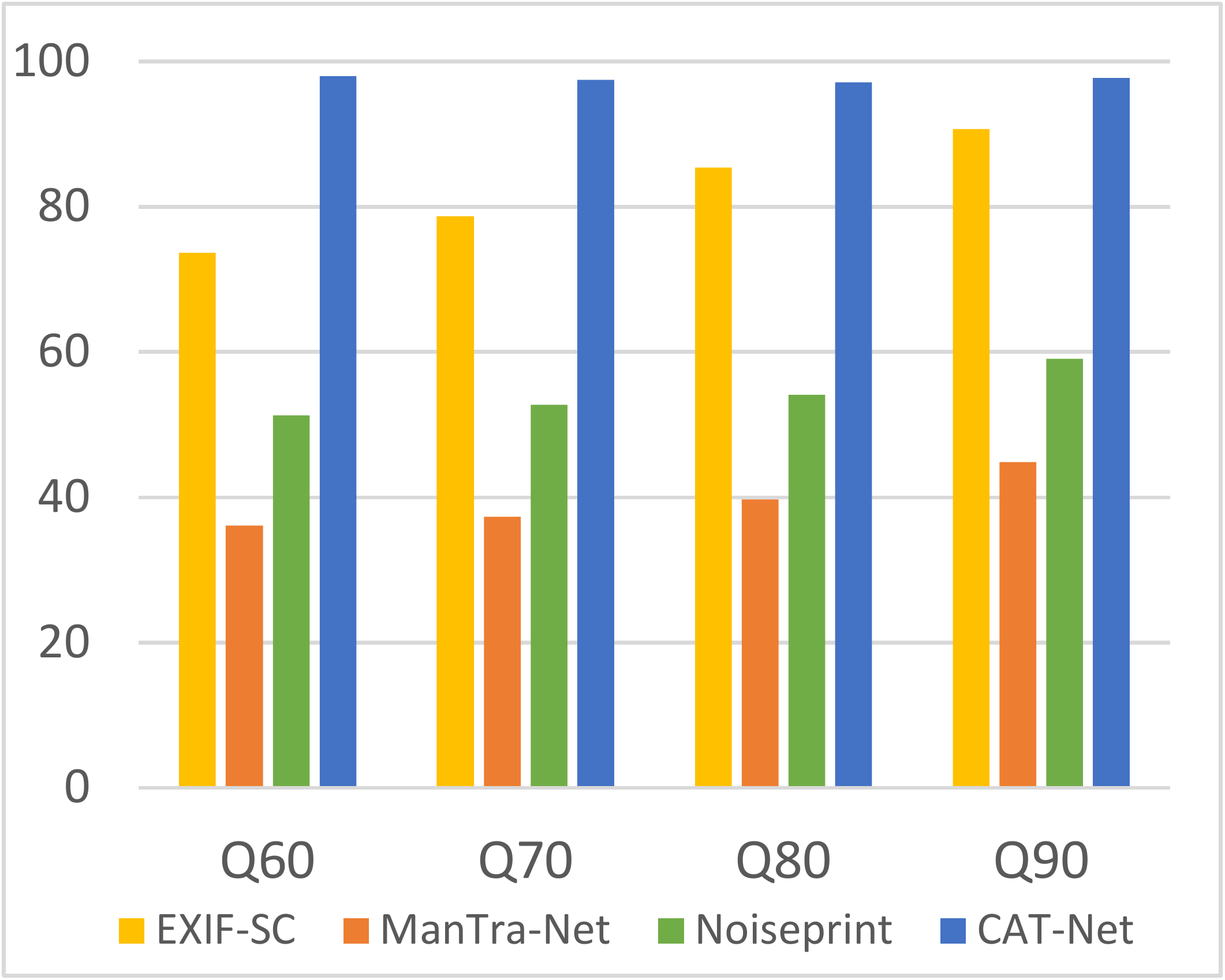}
  }
  \vspace{3mm}
  
  \subfigure[GRIP, p-AP]{
   \includegraphics[width=0.30\linewidth]{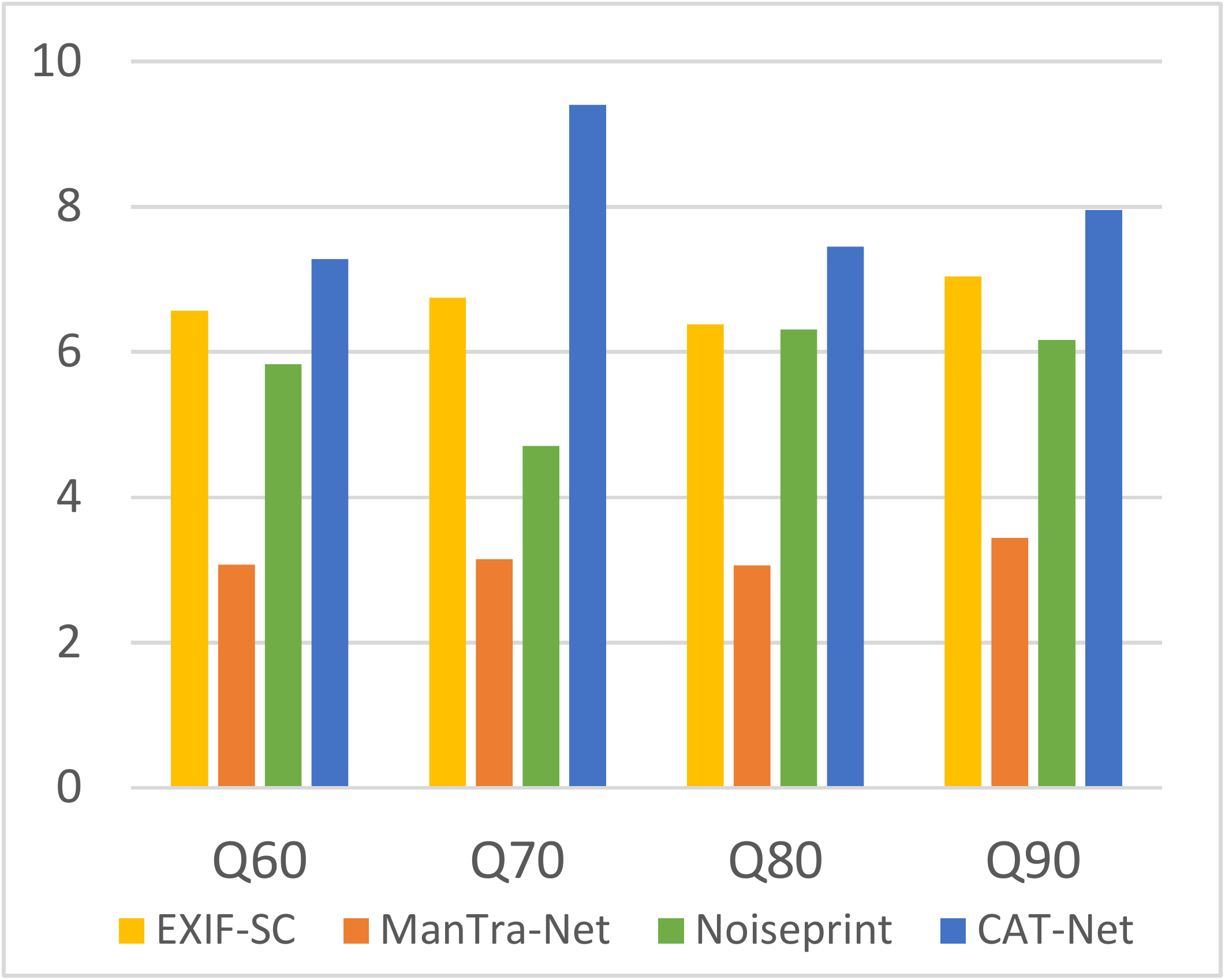}
  }
  \subfigure[CoMoFoD, p-AP]{
   \includegraphics[width=0.30\linewidth]{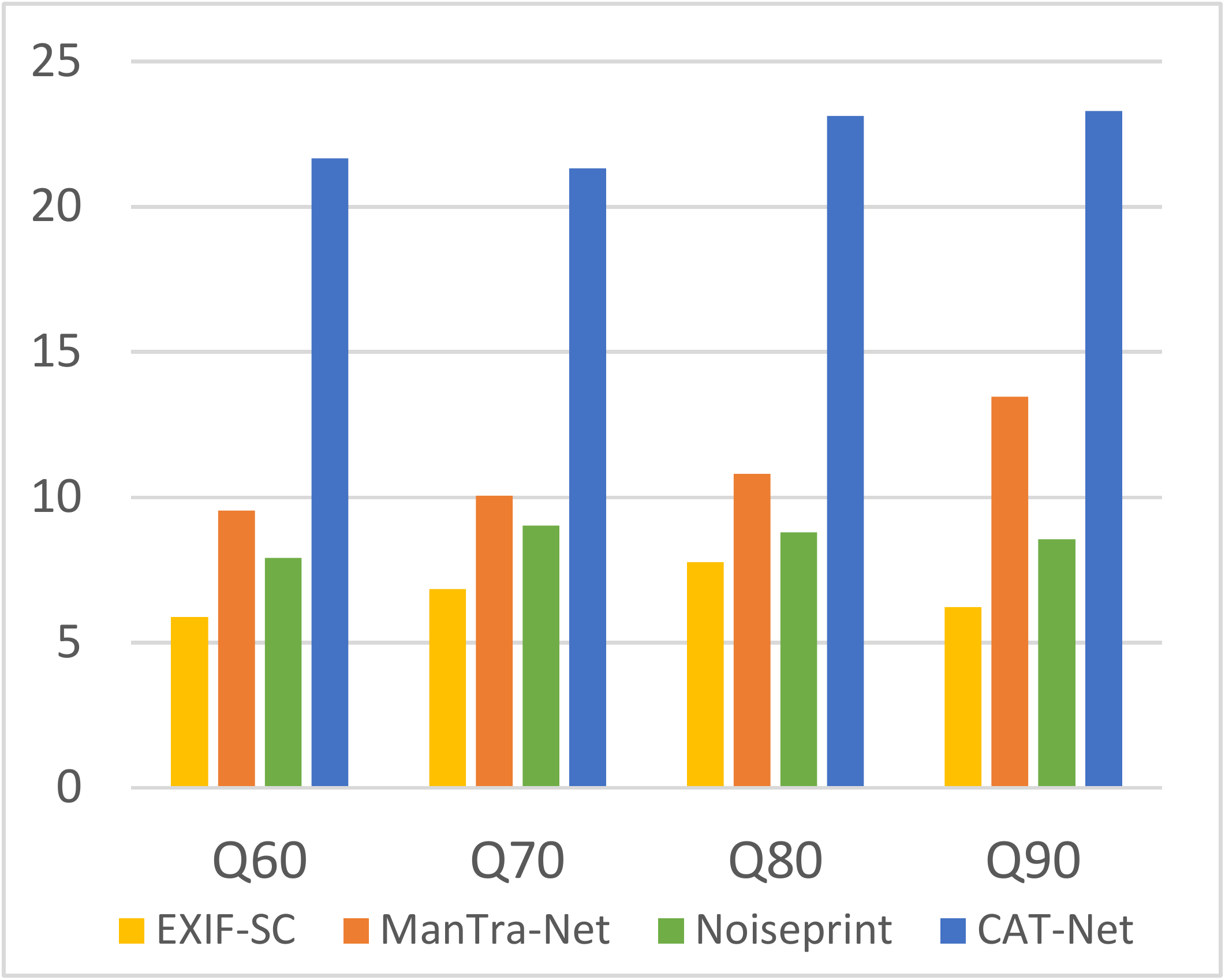}
  }
  \subfigure[COVERAGE, p-AP]{
   \includegraphics[width=0.30\linewidth]{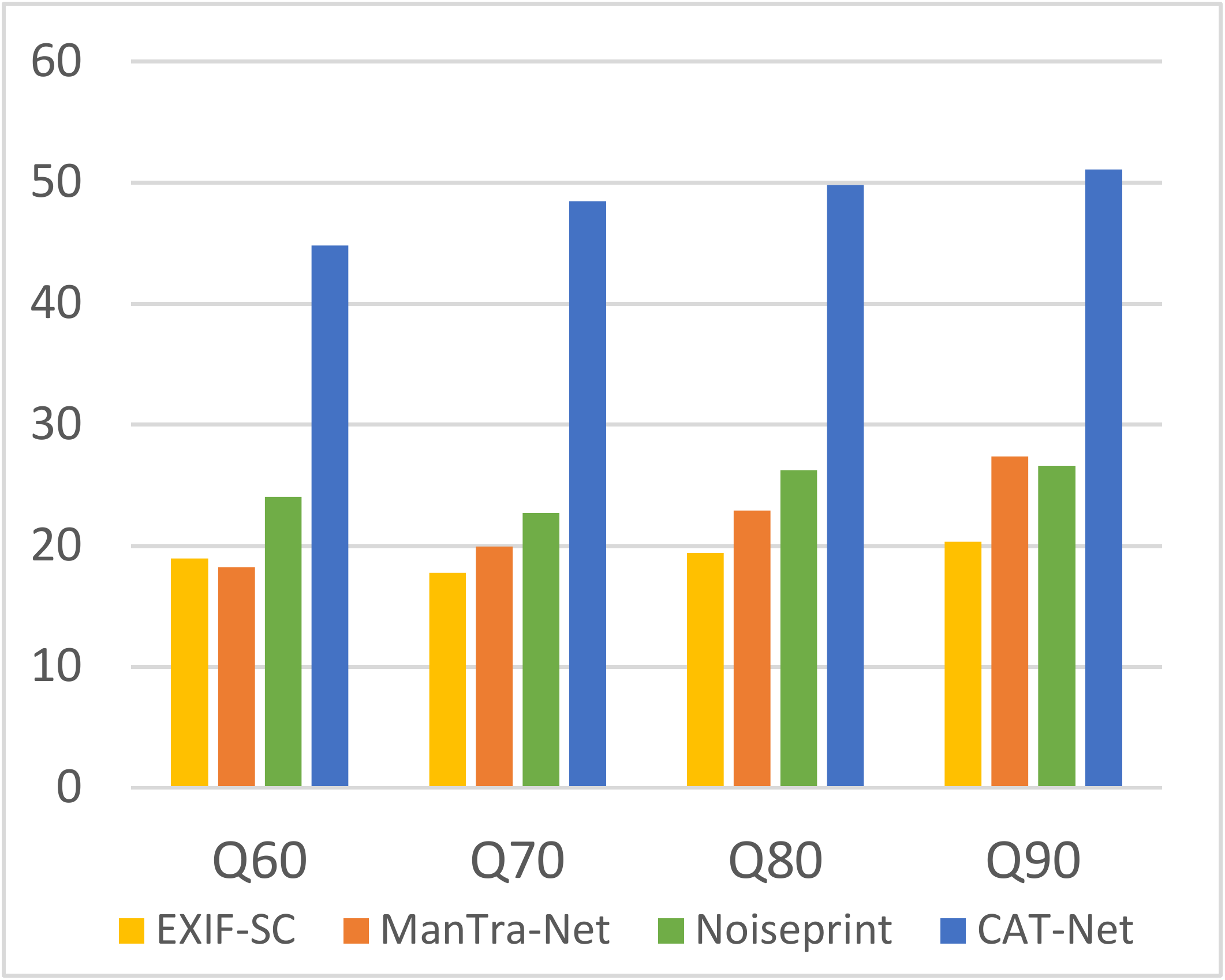}
  }
  \caption{Robustness tests on additional JPEG compression (Sect.~\ref{subsec:cat_rob}).
  CAT-Net achieves the best performance in 16 out of 24 settings in terms of p-F1 score and in 23 out of 24 settings in terms of p-AP. }
  \label{fig:cat_robust}%
  \end{figure*}

  \subsection{Effect of Compression Quality}
  \label{subsec:cat_quant_analysis}
  This subsection analyzes the effect of first and second compression quality. We created forged images similar to Fig.~\ref{fig:cat_problem_b} but with different JPEG compression qualities.
  Recall that the forged image contains the copy-moved object (the cherry blossom tree) and the spliced object (the sign).
  The authentic image (Fig.~\ref{fig:cat_problem_a}) was first compressed using JPEG quality $Q_1$, and another authentic image, the sign, was compressed using JPEG quality 70.
  Then copy-move and splicing manipulation were applied, followed by the second compression using JPEG quality $Q_2$.
  
  Figure~\ref{fig:cat_double_comp} depicts CAT-Net's localization results with diverse compression qualities.
  The smaller $Q_1$ and the larger $Q_2$, the higher the localization performance.
  The diagonal images from the top-left to the bottom-right illustrate the special cases when $Q_1 = Q_2$.
  In these cases, CAT-Net had a lower chance of detecting the copy-moved object because the same quantization produces significantly fewer compression artifacts.
  In contrast, the spliced object was accurately detected even when the same quantization was used in the first and second compression ($Q_2=70$).
  This is mainly because the spliced object was from a different image, unlike the copy-moved object.
  The RGB stream detected that the object had different acquisition artifacts from the other area, so CAT-Net could localize the sign as tampered.
  Furthermore, both objects were hard to detect for images with low second compression qualities ($Q_2 \leq 60$) because the strong final compression extensively destroys the forensic clues.

  \subsection{Robustness Tests on Additional Compression} 
  \label{subsec:cat_rob}
  
  Figure~\ref{fig:cat_robust} illustrates the localization performance when images are JPEG compressed once more. Additional JPEG compression conventionally occurs when the manipulated images are transmitted through the Internet, like posting on social media or sending via messengers. These services often use JPEG compression to reduce storage or bandwidth. Thus, detectors should maintain their performance with additional JPEG compression.
  
  CAT-Net achieves the highest performance in 16 out of 24 settings in terms of p-F1 score and in 23 out of 24 settings in terms of p-AP. Consequently, CAT-Net is robust to additional JPEG compression for various quality factors compared to other neural network approaches and is suitable for detecting real-world forgeries.

  \section{Conclusion}
  \label{sec:cat_conclusion}
  We presented a new approach using image compression artifacts to detect and localize image manipulation. This study is the first to accept DCT coefficients directly into a segmentation network, which was possible due to DCT volume representation and specially chosen neural network components. We also introduced a new pretraining method that uses double JPEG detection. Our neural network approach was the first to use both RGB and DCT domain information for forgery localization. Proposed CAT-Net significantly outperformed state-of-the-art forgery detectors. This study is a starting point for using compression artifacts in deep learning-based image forensics. We hope that many future studies will build upon this idea.
  
  
  \begin{acknowledgements}
  This research was partially supported by Basic Science Research Program through the National Research Foundation of Korea (NRF) funded by the Ministry of Education (2021R1I1A1A01043600).
  \end{acknowledgements}

  %
  \section*{Conflict of interest}
  
  The authors declare that they have no conflict of interest.
  

  \bibliographystyle{spbasic}      
  \bibliography{egbib.bib}   
  
  
  \end{document}